\documentclass[paper]{JHEP3}

\title{Constant terms in threshold resummation and the quark  form factor}
\author{S.\ Friot\\
        Dept. d'Estructura i Constituents de la Mat\`eria, Universitat de Barcelona\\
Diagonal 647, 
E-08028 Barcelona, Catalonia, Spain\\
        E-mail: \email{friot@ecm.ub.es}}

\author{G.\ Grunberg\\
        Centre de Physique Th\'eorique, \'Ecole  
Polytechnique, CNRS,\\
        91128 Palaiseau Cedex, France\\
        E-mail: \email{georges.grunberg@pascal.cpht.polytechnique.fr}}

\abstract{We verify to order $\alpha_s^4$ two previously conjectured relations, valid in four dimensions,
 between constant terms in threshold resummation (for Deep Inelastic Scattering and the Drell-Yan process) and the
second logarithmic derivative of the massless quark form factor. The same relations are checked to all orders in the
large-$\beta_0$ limit; as a by-product a dispersive representation of the form factor is obtained. These relations allow
to compute in a symmetrical way the three-loop resummation coefficients $B_3$ and $D_3$ in terms of the three-loop 
contributions to the virtual diagonal splitting function and to the quark form factor, confirming results obtained in
the literature. }
\preprint{UB-ECM-PF-07-15\\CPHT-RR 046.0607}
\begin{document}


\section{Introduction}
Threshold resummation, namely the resummation to all orders of perturbation theory of the large logarithmic
corrections which arise from the incomplete cancellation of  soft and collinear gluons at the edge of phase
space, is by now a well developed subject \cite{Sterman:1986aj,Catani:1989ne} in perturbative QCD. Large logarithms
are however always accompanied by constant terms, whose contribution may be numerically important. In recent years,
there has been some interest in the way these constant terms organize themselves. In \cite{Grunberg:2006gd}, a
relation between the constant terms and the massless quark  form factor, valid in four dimensions,  was
conjectured in the case of Deep Inelastic Scattering (DIS) and Drell-Yan (DY). In  this paper, we check this
conjecture to order $\alpha_s^4$. In addition, a check to all orders is performed in the large-$n_f$ limit, based on
the dispersive approach.

\noindent The paper is organized as follows. The order  $\alpha_s^4$ check is performed in sections 2 and 3 for DIS 
and DY, respectively. In section 4, we point out that the conjecture, if correct, is valid for the most general
class of resummation procedures. The all-order large-$n_f$  check is performed in section 5, where contact is made with the dispersive approach
\cite{Grunberg:2006gd,Grunberg:2006ky} viewed as a peculiar resummation procedure; as a by-product, a dispersive
representation of the large-$n_f$ quark form factor is obtained. Section 6 contains our conclusions.  More technical issues are dealt with in
three appendices. In Appendix A an original method, based on the Mellin-Barnes representation, is exposed to
compute the large-$N$ (index) behavior of moments of
$+$-distributions, including constant terms. In Appendix B, the large-$N$ scaling behavior of the characteristic
functions which occur in the dispersive approach is derived. In Appendix C the calculation of the massless
one-loop quark form factor  with a finite gluon mass, which gives the characteristic function of the large-$n_f$
quark form factor, is detailed. The   method we use is based on the resummation of the small gluon mass
asymptotic expansion, itself derived through  the Mellin-Barnes representation technique.

\section{Threshold resummation of the physical anomalous dimension (DIS case)}
We recall the standard resummation formula \cite{Sterman:1986aj,Catani:1989ne,Moch:2005ba}. Consider the leading
(twist 2) contribution to the non-singlet structure function $F_2(Q^2,N)$ in Mellin $N$-space
\begin{equation}\label{OPE} F_2(Q^2,N)=O(N,\mu^2)\ {\cal C}(Q^2,N,\mu^2)\ ,\end{equation}
where $O(N,\mu^2)$ is the matrix element, ${\cal C}(Q^2,N,\mu^2)$ the coefficient function, and $\mu^2$  the
factorization scale. At large-$N$ we have, neglecting terms that fall  as $1/N$ (up to logarithms) order by order
in perturbation theory  
\begin{equation}
{\cal C}(Q^2,N,\mu^2)\sim g_{DIS}(Q^2,\mu^2)\
\exp[E_{DIS}(Q^2,N,\mu^2)]\label{resum}\ ,
\end{equation} 
with the Sudakov exponent given by
\begin{equation}
E_{DIS}(Q^2,N,\mu^2)=\int_0^1 dz
\frac{z^{N-1}-1}{1-z}\left[\int_{\mu^2}^{(1-z)Q^2}\frac{dk^2}{k^2}
A\left(a_s(k^2)\right)+B\left(a_s((1-z)Q^2)\right)\right]\ ,\label{exponent}
\end{equation}
where, following the conventions of \cite{Moch:2005ba}, with $a_s\equiv\frac{\alpha_s}{4\pi}$  
\begin{equation}\label{cusp}
A(a_s)=\sum_{i=1}^\infty
A_ia_s^{i}
\end{equation} 
($A$ is the universal ``cusp'' anomalous dimension),
and
\begin{equation}\label{B-stan}
B(a_s)=
\sum_{i=1}^\infty B_i a_s^{i}\ ,
\end{equation}
are the usual Sudakov anomalous dimensions, whereas
\begin{equation}\label{g-stan}
g_{DIS}(Q^2,\mu^2)=1+\sum_{i=1}^\infty  g_i^{DIS}\left(\frac{Q^2}{\mu^2}\right)
a_s^{i}(\mu^2)
\end{equation}
collects the residual constant ($N$-independent) terms not included in $E_{DIS}$.  We note that
$g_{DIS}$ is {\em different} from  $g_0$ as defined in \cite{Moch:2005ba} (which collects {\em all} the constant terms on the right-hand side of eq.(\ref{resum})). Taking the derivative of
eq.(\ref{exponent}) we get
\cite{Forte:2002ni,Gardi:2002xm}
\begin{equation}\label{dE}
{dE_{DIS}(Q^2,N,\mu^2)\over d\ln Q^2}=\int_{0}^1 dz{z^{N-1}-1
 \over
1-z} {\cal J}[(1-z)Q^2]\ ,
\end{equation}
where 
\begin{equation} 
{\cal J}(k^2)= A\left(a_s(k^2)\right)+{dB\left(a_s(k^2)\right)\over d\ln
k^2}\label{standard-S-coupling}\ ,
\end{equation} 
and ${\cal J}$ refers to the ``jet scale'' $(1-z)Q^2$ in eq.(\ref{dE}).
Thus for the ``physical anomalous dimension'' \cite{Grunberg:1982fw,Catani:1996sc} which describes the scaling violation one obtains at large-$N$
\begin{equation}\label{scaling-violation}
{d\ln F_2(Q^2,N)\over d\ln Q^2}={d\ln{\cal C}(Q^2,N,\mu^2)\over d\ln
Q^2}\sim\int_{0}^1 dz{z^{N-1}-1
 \over
1-z} {\cal J}(1-z)Q^2]+ H\left(a_s(Q^2)\right)\ ,
\end{equation}
where
\begin{equation} 
H\left(a_s(Q^2)\right)={d\ln g_{DIS}(Q^2,\mu^2)\over d\ln
Q^2}\label{H-stan}\ .
\end{equation} 
Both the ``Sudakov effective coupling'' 
\begin{equation}\label{A-S}
{\cal J}(k^2)=\sum_{i=1}^\infty
{\cal J}_i a_s^{i}(k^2)
\end{equation}
and the ``leftover'' constant terms function
\begin{equation}\label{H-stan}
H(a_s)=\sum_{i=1}^\infty H_ia_s^{i}
\end{equation}
are renormalization group invariant quantities, given as power series in $a_s$. At the difference of  the usual 
Sudakov ``anomalous dimensions'' $A$ and $B$, they are also renormalization scheme independent quantities. 

\noindent Changing variables to
$k^2=(1-z)Q^2$, eq.(\ref{dE}) becomes identically
\begin{equation}\label{ren-integral-stan}
{dE_{DIS}(Q^2,N,\mu^2)\over d\ln Q^2}=\int_{0}^{Q^2}{dk^2\over k^2}
F_{DIS}\left(\frac{k^2}{Q^2}, N\right)
 {\cal J}(k^2)\ ,
 \end{equation}
with 
\begin{equation}F_{DIS}\left(\frac{k^2}{Q^2}, N\right)=\left(1-\frac{k^2}{Q^2}\right)^{N-1}-1
\label{eq:F-stan}\ .
\end{equation}
It was  shown in \cite{Grunberg:2006hg} that, up to terms which vanish for $N\rightarrow \infty$, we have
\begin{eqnarray}\label{ren-integral-stan2}
{dE_{DIS}(Q^2,N,\mu^2)\over d\ln
Q^2}&\sim&\int_{0}^{Q^2}{dk^2\over k^2} G_{DIS}\left(\frac{N k^2}{Q^2}\right)
 {\cal J}(k^2)\nonumber\\
&\equiv&S_{DIS}(Q^2,N)\ ,
\end{eqnarray}
with
\begin{equation}
G_{DIS}\left(\frac{N k^2}{Q^2}\right)=\exp\left(-\frac{N
k^2}{Q^2}\right)-1\label{G-stan}\ ,
\end{equation} 
where $G_{DIS}(N k^2/Q^2)$ is obtained by taking the $N\rightarrow
\infty$ limit of $F_{DIS}(k^2/Q^2,N)$ with $N k^2/Q^2$ fixed.  Thus we obtain at large-$N$
\begin{equation}\label{scaling-violation1}
{d\ln F_2(Q^2,N)\over d\ln Q^2}\sim S_{DIS}(Q^2,N)+
H\left(a_s(Q^2)\right)\ .
\end{equation} 
An additional simplification is achieved by extending to infinity the upper limit of integration in
eq.(\ref{ren-integral-stan2}), and introducing a suitable UV subtraction term, thus obtaining, up to terms which
vanish for $N\rightarrow\infty$
\begin{eqnarray}\label{ren-integral1} 
S_{DIS}(Q^2,N)&\sim&\int_{0}^{\infty}{dk^2\over k^2} G_{DIS}\left(\frac{N k^2}{Q^2}\right) {\cal J}(k^2)- G_{DIS}(\infty)\int_{Q^2}^{\infty}{dk^2\over k^2} {\cal
J}(k^2)\nonumber\\ &=&
\int_{0}^{\infty}{dk^2\over k^2} G_{DIS}\left(\frac{N k^2}{Q^2}\right) {\cal J}(k^2)+\int_{Q^2}^{\infty}{dk^2\over k^2}
{\cal J}(k^2)\ ,
\end{eqnarray}
 where the (UV finite) combination of the two (separately UV divergent, but IR finite) integrals  on
the right-hand side, when expanded in powers of
$a_s(Q^2)$, contains only
logarithmic and constant terms, and  is free of ${\cal O}\left(\frac{\ln^p N}{N}\right)$ terms
 (at the difference of the left-hand side).
 In the second line, we used that  $G_{DIS}(\infty)=-1$, corresponding to the virtual contribution (the $-1$ on the right-hand side of
eq.(\ref{G-stan})). Thus we have
\begin{equation}
\int_{0}^{\infty}{dk^2\over k^2} G_{DIS}\left(\frac{N k^2}{Q^2}\right) {\cal J}(k^2)
+\int_{Q^2}^{\infty}{dk^2\over k^2}
{\cal J}(k^2)=\sum_{i=1}^\infty \gamma_i(N)\
a_s^{i}(Q^2)\label{eq:S-series}\ ,
\end{equation} 
with ($L\equiv \ln N$)
\begin{eqnarray}
\gamma_1(N)&=&\gamma_{11}L+\gamma_{10}\nonumber\\
\gamma_2(N)&=&\gamma_{22}L^2+\gamma_{21}L+\gamma_{20}\nonumber\\
\gamma_3(N)&=&\gamma_{33}L^3+\gamma_{32}L^2+\gamma_{31}L+\gamma_{30}\label{gammas}\\
&etc.&\nonumber\ .
\end{eqnarray}

\noindent Alternatively, one may  remove the virtual contribution from the Sudakov integral (so that it contains only
 real gluon emission contributions), and merge it together  with the ``leftover'' constant terms,
which yields the equivalent result, in terms of two separately IR divergent (but UV finite) integrals
\begin{equation}\label{ren-integral2}
S_{DIS}(Q^2,N)\sim\int_{0}^{\infty}{dk^2\over k^2} \left[G_{DIS}\left(\frac{N k^2}{Q^2}\right)+1\right] {\cal J}(k^2)- \int_{0}^{Q^2}{dk^2\over k^2} {\cal J}(k^2)\ .
\end{equation}
Using eq.(\ref{ren-integral1}) into eq.(\ref{scaling-violation1}), we end up with the large-$N$ expression
\begin{eqnarray}
{d\ln F_2(Q^2,N)\over d\ln Q^2}&\sim&\int_{0}^{\infty}{dk^2\over k^2} G_{DIS}\left(\frac{N k^2}{Q^2}\right){\cal J}(k^2)\nonumber\\ 
&+&\left[H\left(a_s(Q^2)\right)+\int_{Q^2}^{\infty}{dk^2\over k^2}
{\cal J}(k^2)\right]\label{ren-int-as}\ .
\end{eqnarray} 
If instead one  uses eq.(\ref{ren-integral2}) into eq.(\ref{scaling-violation1}) one gets the equivalent
form
\begin{eqnarray}
{d\ln F_2(Q^2,N)\over d\ln Q^2}&\sim&\int_{0}^{\infty}{dk^2\over k^2} \left[G_{DIS}\left(\frac{N k^2}{Q^2}\right)+1\right]{\cal J}(k^2)\nonumber\\ 
&+&\left[H\left(a_s(Q^2)\right)-\int_{0}^{Q^2}{dk^2\over k^2}
{\cal J}(k^2)\right]\label{ren-int-as-bis}\ .
\end{eqnarray}
Next we observe that the UV (respectively IR) divergences present in the individual integrals in
eq.(\ref{ren-int-as}) (respectively eq.(\ref{ren-int-as-bis}))  disappear after taking one more derivative (which
eliminates the virtual contribution inside the Sudakov integral), namely
\begin{eqnarray}
{d^2\ln F_2(Q^2,N)\over (d\ln Q^2)^2}\sim\int_{0}^{\infty}{dk^2\over k^2}
\dot{G}_{DIS}\left(\frac{N k^2}{Q^2}\right) {\cal J}(k^2)\nonumber\\
+\left[\frac{dH}{d\ln Q^2}-{\cal J}(Q^2)\right]\label{d-scale-viol}\ ,
\end{eqnarray}
where $\dot{G}_{DIS}\equiv- dG_{DIS}/ d\ln k^2$, and
the integral in eq.(\ref{d-scale-viol}) 
\begin{equation}
S^{'}_{DIS}\left(\frac{Q^2}{N}\right)\equiv\int_{0}^{\infty}{dk^2\over k^2}
\dot{G}_{DIS}\left(\frac{N k^2}{Q^2}\right) {\cal J}(k^2)\label{S-dot-exp},
\end{equation}
which depends on the {\em single} variable $Q^2/N$ (the moment space ``jet scale'') is finite.
  In
\cite{Grunberg:2006gd} it was conjectured that the combination
$dH/ d\ln Q^2- {\cal J}(Q^2)$, which represents the ``leftover'' constant terms not included 
in $ S^{'}_{DIS}(Q^2/N)$, is  related to the space-like on-shell electromagnetic
 massless
quark form factor
\cite{Magnea:1990zb} ${\cal F}(Q^2)$ by the identity
\begin{equation}\label{conjecture}
{d^2\ln \left({\cal F}(Q^2)\right)^2 \over
(d\ln Q^2)^2}=\frac{dH}{d\ln Q^2}- {\cal J}(Q^2)\ .
\end{equation}
If this conjecture is correct, the second
line of eq.(\ref{ren-int-as-bis}), which involves an IR divergent integral,  could be formally identified  to the
first logarithmic derivative of the square of the quark form factor, an IR divergent
 quantity:
 \begin{equation}\label{conjecture-0}
 {d\ln \left({\cal
F}(Q^2)\right)^2 \over d\ln Q^2}=H\left(a_s(Q^2)\right)-\int_{0}^{Q^2}{dk^2\over k^2}
{\cal J}(k^2)\ .
\end{equation}
In this section, we show that  eq.(\ref{conjecture})
can be checked to order $a_s^4$, using results in the litterature 
\cite{Moch:2005ba,Moch:2005id,Moch:2004pa}.

Let us first consider the left-hand side of eq.(\ref{conjecture}).
We begin from the evolution equation satisfied \cite{Magnea:1990zb} by the form factor in
$D=4-\epsilon$ dimensions  (after multiplication by a factor of
$2$):
\begin{equation}\label{evolution}
{d\ln \left({\cal F}(Q^2,\epsilon)\right)^2 \over
d\ln Q^2}=K\left(a_s(\mu^2),\epsilon\right)+G\left(\frac{Q^2}{\mu^2},a_s(\mu^2),\epsilon\right)\ ,
\end{equation}
where $K\left(a_s(\mu^2),\epsilon\right)$ is a counterterm function which contains only poles in $1/\epsilon$,
and is independent of
$Q^2$, while 
$G\left(\frac{Q^2}{\mu^2},a_s(\mu^2),\epsilon\right)$ is {\em finite} in four dimensions. Taking a second 
derivative and letting
$\epsilon=0$  yields 
\begin{equation}
\frac{d^2\ln\left(\mathcal{F}(Q^2)\right)^2}{(d\ln Q^2)^2}=\frac{d}{d\ln
Q^2}G\left(\frac{Q^2}{\mu^2},a_s(\mu^2)\right)=-\mu^2\frac{\partial}{\partial
\mu^2}G\left(\frac{Q^2}{\mu^2},a_s(\mu^2)\right)\ .
\end{equation}
We now use the renormalization group equation satisfied \cite{Magnea:1990zb} by the $G$ function  (at
$\epsilon=0$)
\begin{equation}\left(\mu^2\frac{\partial}{\partial
\mu^2}+
\beta\left(a_s\right)\frac{\partial}{\partial
a_s}\right)G\left(\frac{Q^2}{\mu^2},a_s\right)=A\left(a_s\right)
\end{equation}
to find
\begin{equation}
\frac{d^2\ln\left(\mathcal{F}(Q^2)\right)^2}{(d\ln Q^2)^2}=\beta\left(a_s(\mu^2)\right)\frac{\partial}{\partial
a_s}G\left(\frac{Q^2}{\mu^2},a_s(\mu^2)\right)-A\left(a_s(\mu^2)\right)\ ,
\end{equation}
where  the beta function is given by the series
\begin{equation}
\beta(a_s)=-\sum_{i=0}^\infty\beta_ia_s^{i+2}\label{beta}\ .
\end{equation}
Now, since $\frac{d^2\ln\left(\mathcal{F}(Q^2)\right)^2}{d(\ln Q^2)^2}$ is a renormalisation group invariant
quantity, we can set $\mu=Q$, and get
\begin{equation}\label{left}
\frac{d^2\ln\left(\mathcal{F}(Q^2)\right)^2}{(d\ln Q^2)^2}=\beta\left(a_s(Q^2)\right)\frac{\partial}{\partial
a_s}G\left(1,a_s(Q^2)\right)-A\left(a_s(Q^2)\right)\ .
\end{equation}
This relation is useful because it will give us the possibility to use the expressions for $G$  quoted 
 in \cite{Moch:2005id} for $\mu=Q$.
Rewriting eq.(\ref{conjecture}) as
\begin{equation}\label{conjecture1} 
{\cal J}(Q^2)=
-\frac{d^2\ln\left(\mathcal{F}(Q^2)\right)^2}{(d\ln Q^2)^2}+\frac{dH}{d\ln Q^2}
\end{equation}
and using
\begin{equation}\label{dH_{new}}\frac{dH}{d\ln Q^2}=
\beta\left(a_s(Q^2)\right)\frac{\partial}{\partial
a_s}H\left(a_s(Q^2)\right)
\end{equation}
the relation to be checked becomes
\begin{equation}\label{A-dB}
 {\cal J}(Q^2)=A\left(a_s(Q^2)\right)+\beta\left(a_s(Q^2)\right)\frac{\partial}{\partial
a_s}\left[-G\left(1,a_s(Q^2)\right)+H\left(a_s(Q^2)\right)\right]\ .
\end{equation}
Therefore, comparing  with eq.(\ref{standard-S-coupling}),
we  have to check that   
\begin{equation}\label{B-H}
B(a_s)=-G(1,a_s)+H(a_s)\ .
\end{equation}
This is essentially a check of the ``non-conformal'' part \footnote{Eq.(\ref{B-H}) also shows
 that the combination $G(1,a_s)+B(a_s)$ is renormalization scheme invariant.} of ${\cal J}$.
 We note that a contribution at order $a_s^i$ to $B$ implies a contribution at order $a_s^{i+1}$ to
 $ {\cal J}$ in eq.(\ref{standard-S-coupling}). In particular, since $B$ starts at order $a_s$, it will contribute to 
${\cal J}$ only starting at order  $a_s^2$, and the order $a_s$ contribution will be entirely provided by 
the cusp anomalous dimension $A$, yielding ${\cal J}_1=A_1$.  Moreover,  a check of eq.(\ref{B-H})
to order
$a_s^i$  implies a check of eq.(\ref{A-dB}) to order
$a_s^{i+1}$. Since the cusp anomalous dimension $A(a_s)$ does not appear explicitly in eq.(\ref{B-stan}), this
 observation implies that we shall be able to check eq.(\ref{conjecture}) to order $a_s^4$, despite
the fact that $A(a_s)$ is only known
to order\footnote{Actually, as we shall see, even $A_3$ is not needed up to this order.}
$a_s^3$. Following the conventions of \cite{Moch:2005id} for the $G$  function, 
\begin{equation}\label{G}
G(1,a_s)=\sum_{i=1}^\infty G_ia_s^{i}\ ,
\end{equation}
  we therefore see that the proof of eq.(\ref{conjecture}) amounts
 to show that
\begin{equation}\label{B-H-bis}
G(1,a_s)=H(a_s)-B(a_s)\ ,
\end{equation}
i.e., for all $i\geq 1$,
\begin{equation}
G_{i}= H_{i}-B_{i}\label{check}\ .
\end{equation}

Let us come into the details of the proof for $i\leq 3$. The $G_i$'s and $B_i$'s have been computed
in the litterature up to $i=3$ but for the $H_i$'s a bit of work is still needed.
We have the following relation to determine  $H_i$:
\begin{equation}\label{H}
H\left(a_s(Q^2)\right)+ C_{DIS}\left(a_s(Q^2)\right)=\frac{d}{d\ln Q^2}\ln
g_0^{DIS}\left(\frac{Q^2}{\mu^2},a_s(\mu^2)\right)\ ,
\end{equation}
where $g_0^{DIS}$ is a function\footnote{$g_0^{DIS}(1,a_s)$ is denoted $g_0(a_s)$ in \cite{Moch:2005ba}.} which collects {\em all} the
constant  terms on the right-hand side of eq.(\ref{resum}), and is thus  {\em
different}
 from
$g_{DIS}$, whereas
$C_{DIS}$ collects the constant terms included in the Sudakov integrals on the right-hand side of
eq.(\ref{ren-integral-stan}) or (\ref{ren-integral-stan2}).
Eq.(\ref{H}) simply expresses the fact that the constant terms on the right-hand side of
eq.(\ref{scaling-violation1}) are  the sum of the constant terms originating from $H$ and those included 
in the
Sudakov integral $S_{DIS}(Q^2,N)$.
For $\mu=Q$ we have
\begin{equation}
g_0^{DIS}(1,a_s)=1+\sum_{i=1}^\infty g_{0i}^{DIS}\ a_s^{i}\label{g0i}\ ,
\end{equation}
where the $g_{0i}^{DIS}$'s are known \cite{Moch:2005ba} up to $i=3$ and we show below that one can obtain $\frac{d}{d\ln Q^2}\ln
g_0^{DIS}\left(\frac{Q^2}{\mu^2},a_s(\mu^2)\right)$ from $g_0^{DIS}(1,a_s)$. On the other hand
$C_{DIS}$ is given by the series (see eq.(\ref{gammas}))
\begin{equation}
C_{DIS}(a_s)=\sum_{i=1}^\infty \gamma_{i0}\
a_s^{i}\label{gamma0-series}\ .
\end{equation}

\noindent We begin with the calculation of  the right-hand side of eq.(\ref{H}). We have
\begin{equation}\label{dg0}
\frac{d}{d\ln Q^2}\ln g_0^{DIS}\left(\frac{Q^2}{\mu^2},a_s(\mu^2)\right)=-\mu^2\frac{\partial}{\partial \mu^2}\ln
g_0^{DIS}\left(\frac{Q^2}{\mu^2},a_s(\mu^2)\right)\ .
\end{equation}
We can then use eq.(3.10) of \cite{Moch:2004pa} to obtain the renormalization group equation satisfied by
 $g_0^{DIS}$:
\begin{equation}
\left[\mu^2\frac{\partial}{\partial \mu^2}+\beta\left(a_s(\mu^2)\right)\frac{\partial}{\partial
a_s}\right]\ln
g_0^{DIS}\left(\frac{Q^2}{\mu^2},a_s(\mu^2)\right)=A\left(a_s(\mu^2)\right)\gamma_E-B_{\delta}\left(a_s(\mu^2)\right)
\label{RG-g0}\ ,
\end{equation}
where $B_{\delta}$  is the coefficient of $\delta(1-x)$ in the non-singlet splitting function.  Its expansion
in powers of
$a_s$ 
\begin{equation}
B_{\delta}(a_s)=\sum_{i=1}^\infty B_{i}^{\delta} a_s^{i}\label{Bi-delta}
\end{equation}
is known up to $i=3$ and can be found in
 \cite{Moch:2004pa}, \cite{Idilbi:2006dg}. We thus obtain
\begin{eqnarray}
\frac{d}{d\ln Q^2}\ln g_0^{DIS}\left(\frac{Q^2}{\mu^2},a_s(\mu^2)\right)&=&
-[A\left(a_s(\mu^2)\right)\gamma_E-B_{\delta}\left(a_s(\mu^2)\right)]\nonumber\\
&+&\beta\left(a_s(\mu^2)\right)
\frac{\partial}{\partial
a_s}\ln g_0^{DIS}\left(\frac{Q^2}{\mu^2},a_s(\mu^2)\right)\label{dg0}\ .
\end{eqnarray}
Now, since $\frac{d}{d\ln Q^2}\ln g_0^{DIS}\left(\frac{Q^2}{\mu^2},a_s(\mu^2)\right)$ is  a renormalisation group
 invariant quantity, we can set $\mu^2=Q^2$ on the right-hand side of eq.(\ref{dg0}) to get
\begin{eqnarray}
\frac{d}{d\ln Q^2}\ln g_0^{DIS}\left(\frac{Q^2}{\mu^2},a_s(\mu^2)\right)&=&
[B_{\delta}\left(a_s(Q^2)\right)-A\left(a_s(Q^2)\right)\gamma_E]\nonumber\\
&+&\beta\left(a_s(Q^2)\right)\frac{\partial}{\partial
a_s}\ln g_0^{DIS}\left(1,a_s(Q^2)\right)\label{dg0bis}\ .
\end{eqnarray}

\noindent We then go on by computing  the series coefficients in
eq.(\ref{gamma0-series}) up to order
$a_s^3$. Since ${\cal J}(k^2)$ is a renormalization group invariant effective charge,  we have
the well-known renormalization group logarithmic structure (see e.g. \cite{Grunberg:1982fw}), expanding in powers of
$a_s(Q^2)$
\begin{eqnarray}
{\cal J}(k^2)&=&{\cal J}_1
a_s(Q^2)+\left[-\beta_0
{\cal J}_1\ln\left(\frac{k^2}{Q^2}\right)+{\cal J}_2\right]a_s^2(Q^2)\label{As-logs}
\\ &+&\left[\beta_0^2
{\cal J}_1\ln^2\left(\frac{k^2}{Q^2}\right)-\left(\beta_1 {\cal J}_1+2\beta_0
{\cal J}_2\right)\ln\left(\frac{k^2}{Q^2}\right) +{\cal J}_3\right]a_s^3(Q^2)+...\ .\nonumber
\end{eqnarray}
Moreover eq.(\ref{standard-S-coupling})   gives:
\begin{equation}
{\cal J}_1= A_1\label{A1-stan}\ ,
\end{equation}
\begin{equation}
{\cal J}_2= A_2-\beta_0 B_1\label{A2-stan}
\end{equation}
and
\begin{equation}
{\cal J}_3= A_3-\beta_1B_1-2\beta_0 B_2\label{A3-stan}\ .
\end{equation}
Then the calculation of the coefficients $\gamma_{i0}$ in eq.(\ref{gamma0-series}) up to $i=3$ is reduced
to the evaluation of the constant terms $c_p$ in the $N\rightarrow\infty$ asymptotic expansion of the
 integrals (for $p=0, 1, 2$):
\begin{equation}
I_p(N)\equiv
\int_{0}^{Q^2}{dk^2\over k^2}\Big[
\exp\left(-\frac{N
k^2}{Q^2}\right)-1\Big]\ln^p\left(\frac{k^2}{Q^2}\right)\label{I-p}\ ,
\end{equation}
because, as we have seen, $C_{DIS}$ collects the constant terms of the right-hand side of eq.(\ref{ren-integral-stan2}).
The $c_p$ can be obtained from standard results in the litterature (see e.g. \cite{Moch:2005ba}), since changing
variable to
$k^2/Q^2=1-z$, one gets 
\begin{equation}
I_p(N)=
\int_{0}^1dz
{\exp[-N(1-z)]-1\over 1-z}\ln^p\left(1-z\right)\label{I-p1}\ ,
\end{equation}
which was shown \cite{Catani:1990rr} to have the same large-$N$ expansion (up to 
 terms which vanish for $N\rightarrow\infty$) as 
${\bar I}_p(N)\equiv
\int_{0}^1dz
{z^{N-1}-1\over 1-z}\ln^p\left(1-z\right)$.
One thus gets, for $p=0, 1, 2$:
\begin{equation}
c_0=-\gamma_E\ ,
\label{c0}
\end{equation}
\begin{equation}
c_1=\frac{1}{2}\left(\gamma_E^2+\frac{\pi^2}{6}\right)\ 
\label{c1}
\end{equation}
and
\begin{equation}
c_2=-\frac{1}{3}\left(\gamma_E^3+\gamma_E \frac{\pi^2}{2}+2\zeta_3\right)\ .
\label{c2}
\end{equation}
A novel derivation of these results is presented in Appendix A.
Using eq.(\ref{As-logs}), we then find the series in
eq.(\ref{gamma0-series}) to be given by
\begin{eqnarray}
C_{DIS}(a_s)&=&{\cal J}_1 c_0
a_s+(-\beta_0 {\cal J}_1 c_1+{\cal J}_2 c_0)a_s^2\nonumber\\
&+&[\beta_0^2 {\cal J}_1 c_2-(\beta_1 {\cal J}_1+2\beta_0 {\cal J}_2)c_1
+{\cal J}_3 c_0]a_s^3+...\label{gamma0}\ .
\end{eqnarray}

\noindent Now eqs.(\ref{H}) and (\ref{dg0bis}) yield
\begin{equation}
H(a_s)=B_{\delta}(a_s)-[ C_{DIS}(a_s)+A(a_s)\gamma_E]+\beta(a_s)
\frac{\partial}{\partial
a_s}\ln g_0^{DIS}(1,a_s)\label{H-bis}\ .
\end{equation}
For $i=1$ one thus gets
\begin{equation}
H_1=B_1^{\delta}-[A_1\gamma_E+{\cal J}_1 c_0]=B_1^{\delta}\label{h1}\ ,
\end{equation}
where we used eq.(\ref{A1-stan}).

\noindent For $i=2$ one gets
\begin{eqnarray}\nonumber
H_2&=&B_2^{\delta}-\left[A_2\gamma_E+({\cal J}_2 c_0-\beta_0 {\cal J}_1 c_1)\right]
-\beta_0 g_{01}^{DIS}\\
\label{h2}
&=&B_2^{\delta}-\beta_0\left[\gamma_E B_1-A_1 c_1+g_{01}^{DIS}\right]\ ,
\end{eqnarray}
where we used eq.(\ref{A2-stan}). Thus
\begin{equation}
H_2=B_2^{\delta}+C_F\beta_0(9+4\zeta_2)\nonumber\label{h2-bis}\ ,
\end{equation}
where we used \cite{Moch:2005ba}
\begin{equation}\label{A1}
A_1=4C_F\ ,
\end{equation}
\begin{equation}\label{B1-stan}
B_1=-3C_F
\end{equation}
and 
\begin{equation}
g_{01}^{DIS}=C_F(-9+3\gamma_E+2\gamma_E^2-2\zeta_2)  \label{g01}\ .
\end{equation}

\noindent Finally, for $i=3$  we find
\begin{eqnarray}\nonumber
H_3&=&B_3^{\delta}-\left[A_3\gamma_E+
\left({\cal J}_3 c_0-(\beta_1 {\cal J}_1+2\beta_0 {\cal J}_2)c_1+\beta_0^2 {\cal J}_1 c_2
\right)\right]\nonumber\\
&-&\left[\beta_1g_{01}^{DIS}+\beta_0(2g_{02}^{DIS}-(g_{01}^{DIS})^2)\right]\nonumber\\
\label{h3}
&=&B_3^{\delta}-\beta_1\left[\gamma_E B_1-
A_1 c_1+g_{01}^{DIS}
\right]-\beta_0^2\left[2 c_1 B_1+A_1 c_2\right]\\
&-&\beta_0\left[2\gamma_E B_2-2 A_2 c_1+2g_{02}^{DIS}-(g_{01}^{DIS})^2\right]
\nonumber\ ,
\end{eqnarray}
where we used eq.(\ref{A3-stan}). We note that only $A_j$ with $j<i$ contributes to $H_i$, and that the
coefficient of the $\beta_1$ term in eq.(\ref{h3})  is the same as that of the $\beta_0$ term in
eq.(\ref{h2}) (the general structure underlying these observations will be displayed below). Eq.(\ref{h3}) gives
\begin{eqnarray}
H_3&=&B_3^{\delta}+4 C_F\beta_1\left(\frac{\frac{9}{4}+\pi^2}{6}\right)+4C_F\beta_0^2\left(
\frac{3}{4}\gamma_E^2+\frac{1}{3}\gamma_E^3+\frac{\pi^2}{8}
+\gamma_E
\frac{\pi^2}{6}+\frac{2}{3}\zeta_3\right)\nonumber\\
&+&4C_F\beta_0\left[C_F\left(-\frac{7}{16}-\frac{25}{8}\pi^2+\frac{\pi^4}{60}+33\zeta_3\right)\label{h3bis}\right.\\
&-& C_A\left(-\frac{5465}{144}
+\frac{11}{4}\gamma_E^2+\frac{11}{9}\gamma_E^3-\frac{469}{72}\pi^2+\frac{11}{18}\gamma_E\pi^2+\frac{71}{360}\pi^4
+\frac{232}{9}\zeta_3\right)\nonumber\\
&+&n_f\left.\left(\frac{-457}{72}+\frac{1}{2}\gamma_E^2+\frac{2}{9}\gamma_E^3-\frac{35}{36}\pi^2+\frac{1}{9}\gamma_E\pi^2
-\frac{2}{9}\zeta_3\right)\right]\nonumber\ ,
\end{eqnarray}
where we used \cite{Moch:2005ba}
\begin{equation}\label{A2} 
A_2=C_F\left[\left(\frac{268}{9}-8\zeta_2\right)C_A-\frac{40}{9}n_f\right]\ ,
\end{equation}
\begin{equation}\label{B2}
B_2=C_FC_A\left(-{3155\over54}+{44\over3}\zeta_2+40\zeta_3
\right)+C_Fn_f\left({247\over27}-{8\over3}\zeta_2\right)-C_F^2\left({3\over2}-12\zeta_2+24\zeta_3\right)
\end{equation}
and
\begin{eqnarray}
g_{02}^{DIS}&=&C_F^2\left(\frac{331}{8}-\frac{51}{2}\gamma_E-\frac{27}{2}\gamma_E^2+6\gamma_E^3+2\gamma_E^4
+\frac{111}{2}\zeta_2\right.\nonumber\\
&&\ \ \ \ \ \ \ \ -\left.18\gamma_E\zeta_2-4\gamma_E^2\zeta_2-66\zeta_3+24\gamma_E\zeta_3+\frac{4}{5}\zeta_2^2\right)
\nonumber\\
&+&C_FC_A\left(-\frac{5465}{72}+\frac{3155}{54}\gamma_E+\frac{367}{18}\gamma_E^2+\frac{22}{9}\gamma_E^3
-\frac{1139}{18}\zeta_2\right.\nonumber\\
&&\ \ \ \ \ \ \ \ \ \ \ -\left.\frac{22}{3}\gamma_E\zeta_2-4\gamma_E^2\zeta_2+\frac{464}{9}\zeta_3-40\gamma_E\zeta_3
+\frac{51}{5}\zeta_2^2\right)\nonumber\\
&+&C_Fn_f\left(\frac{457}{36}-\frac{247}{27}\gamma_E-\frac{29}{9}\gamma_E^2-\frac{4}{9}\gamma_E^3
+\frac{85}{9}\zeta_2+\frac{4}{3}\gamma_E\zeta_2+\frac{4}{9}\zeta_3\right)\label{g02}\ .
\end{eqnarray}
Substituting $n_f=-\frac{3}{2}\beta_0+\frac{11}{2}C_A$  in the coefficient of the $\beta_0$ term in
eq.(\ref{h3bis}) yields a further simplification:
\begin{eqnarray}
H_3&=&B_3^{\delta}+C_F\beta_1\left(9+4\zeta_2\right)+C_F\beta_0^2\left(
\frac{457}{12}
+38\zeta_2+4\zeta_3\right)\label{h3ter}\\
&+&C_F\beta_0\left[C_F\left(-\frac{7}{4}-75\zeta_2+132\zeta_3+\frac{12}{5}\zeta_2^2\right)
+ C_A\left(\frac{73}{6}
+28\zeta_2-108\zeta_3-\frac{142}{5}\zeta_2^2
\right)\right]\nonumber\ .
\end{eqnarray} 
We note that  all $\gamma_E$ terms, which arise entirely from the Sudakov
integral (see eq.(\ref{gamma0})), have  cancelled in the $H_i$'s.

We now  have all ingredients to check eq.(\ref{check}) for $i\leq 3$ and verify  eq.(\ref{conjecture}) order by order in  $a_s$, up to order $a_s^4$.
Since  \cite{Moch:2004pa,Idilbi:2006dg}
\begin{equation}\label{B1-delta}
B_1^{\delta}=3C_F\ ,
\end{equation}
we get from eq.(\ref{h1})
\begin{equation}
H_1=3C_F\ .\label{h1-bis}
\end{equation}
Moreover we have \cite{Moch:2005id}
\begin{equation}\label{G1}
G_1=6C_F\ .
\end{equation}
Thus we find, using eq.(\ref{B1-stan}), that
\begin{equation}
G_1=H_1-B_1\ ,\label{check1}
\end{equation}
checking eq.(\ref{check})  for $i=1$.

\noindent We next consider the case $i=2$. 
Since  \cite{Moch:2004pa,Idilbi:2006dg}
\begin{equation}
B_2^{\delta}= C_FC_A \left({17\over6}+{44\over3} \zeta_2 -12\zeta_3\right)
-C_Fn_f \left({1 \over3} + {8 \over 3} \zeta_2 \right)
+ C_F^2\left( {3 \over 2} - 12\zeta_2+24 \zeta_3 \right) \label{B2-delta}
\end{equation}
and
\begin{equation}
\beta_0={11\over3}C_A-{2\over3}n_f  \label{beta-0}\ ,
\end{equation}
eq.(\ref{h2-bis}) gives
\begin{equation}
H_2=C_F
C_A\left({215\over6}+{88\zeta_2\over3}-12\zeta_3\right)-C_F n_f\left(\frac{19}{3}+{16\over3}\zeta_2\right)+
C_F^2\left({3
\over 2} - 12\zeta_2+24 \zeta_3 \right)\label{h2-ter}\ .
\end{equation}
Moreover we have \cite{Moch:2005id}
\begin{equation}\label{G2}
G_2= C_F C_A \left({2545 \over 27} + {44 \over 3} \zeta_2 - 52  \zeta_3\right)
- C_F n_f  \left( {418 \over 27} + {8 \over 3}  \zeta_2\right)+C_F^2 (3 - 24 \zeta_2 + 48 \zeta_3)\ .
\end{equation}
We thus find, using eq.(\ref{B2})
\begin{equation}
G_2=H_2-B_2\ ,
\end{equation}
checking  eq.(\ref{check})  for $i=2$. We also note the simple relations between the $C_F^2$ terms in eq.(\ref{B2}),
(\ref{B2-delta}), (\ref{h2-ter}) and (\ref{G2}). 

\noindent Let us finally consider the case $i=3$.
Since  \cite{Moch:2004pa,Idilbi:2006dg}
\begin{eqnarray}\nonumber
B_3^{\delta}&=&C_F^3 \left({29\over2}+18 \zeta_2+68\zeta_3 +{288\over5}
\zeta_2^2-32\zeta_2\zeta_3-240\zeta_5\right) \\
&+&C_F^2C_A \left({151\over4}-{410\over3} \zeta_2+\frac{844}{3}\zeta_3 -{988\over15}
\zeta_2^2+16\zeta_2\zeta_3+120\zeta_5\right)\nonumber\\
&-&C_FC_A^2 \left({1657\over36}-{4496\over27} \zeta_2+\frac{1552}{9}\zeta_3 +2
\zeta_2^2-40\zeta_5\right)\nonumber\\
&-&C_Fn_f^2 \left({17\over9}-{80\over27} \zeta_2 +\frac{16}{9}\zeta_3\right)
-C_F^2n_f \left(23-{20\over3} \zeta_2+\frac{136}{3}\zeta_3 -{232\over15} \zeta_2^2\right)\nonumber\\
&+&C_FC_An_f \left(20-{1336\over27} \zeta_2+\frac{200}{9}\zeta_3 +{4\over5} \zeta_2^2\right)\label{B3-delta}
\end{eqnarray}
and
\begin{equation}
\beta_1={34\over3}C_A^2-{2}C_Fn_f-\frac{10}{3}C_An_f  \ ,
\end{equation}
we find, using  eq.(\ref{h3bis}),
\begin{eqnarray}\nonumber
H_3&=&C_F^3\left(\frac{29}{2}+18\zeta_2+68\zeta_3+\frac{288}{5}\zeta_2^2-32\zeta_2\zeta_3-240\zeta_5\right)\\
&+&C_F^2C_A \left(\frac{94}{3}-\frac{1235}{3}\zeta_2+\frac{2296}{3}\zeta_3-\frac{856}{15}\zeta_2^2+16\zeta_2\zeta_3
+120\zeta_5\right)\nonumber\\
&+&C_F
 C_A^2
\left(\frac{16540}{27}+\frac{22286}{27}\zeta_2-\frac{1544}{3}\zeta_3-\frac{1592}{15}\zeta_2^2+40\zeta_5\right)
\nonumber\\
&+&C_F^2 n_f\left(-\frac{239}{6}+\frac{146}{3}\zeta_2-\frac{400}{3}\zeta_3+\frac{208}{15}\zeta_2^2\right)+ C_F
n_f^2\left(\frac{406}{27}+\frac{536}{27}\zeta_2\right)
\nonumber\\
&+&C_F C_A  n_f\left(-\frac{5516}{27}-\frac{7216}{27}\zeta_2+\frac{224}{3}\zeta_3+\frac{296}{15}\zeta_2^2\right)
\label{h3ter}\ .
\end{eqnarray}
 Now we have  \cite{Moch:2005ba}
\begin{eqnarray}\nonumber
B_3&=& C_F^3 \left(-{29\over2}-18\zeta_2-68\zeta_3-{288\over5} \zeta_2^2+32\zeta_2\zeta_3+240\zeta_5\right)\\
&+&C_F^2C_A \left(-46+287\zeta_2-\frac{712}{3}\zeta_3-{272\over5} \zeta_2^2 -16\zeta_2\zeta_3-120\zeta_5\right)
\nonumber\\
&+&C_FC_A^2 \left(-\frac{599375}{729}+\frac{32126}{81}\zeta_2+\frac{21032}{27}\zeta_3-{652\over15} \zeta_2^2
-\frac{176}{3}\zeta_2\zeta_3-232\zeta_5\right)
\nonumber\\
&+&C_F^2n_f \left({5501\over54}-50 \zeta_2 +\frac{32}{9}\zeta_3\right)+C_Fn_f^2 \left(-{8714\over729}+{232\over27}
\zeta_2 -{32\over27} \zeta_3\right)\nonumber\\ 
&+&C_FC_An_f \left({160906\over729}-{9920\over81} \zeta_2-{776\over9}
\zeta_3 +\frac{208}{15}\zeta_2^2\right)\label{B3-stan}
\end{eqnarray}
and \cite{Moch:2005id}
\begin{eqnarray}
G_3&=&C_F^3 \left(29+36\zeta_2+ 136
\zeta_3 +\frac{576}{5}\zeta_2^2-64\zeta_2\zeta_3-480\zeta_5\right)\nonumber\\
 &+& C_F^2 C_A \left({232 \over 3} - {2096 \over 3}
\zeta_2+\frac{3008}{3} 
\zeta_3 -\frac{8}{3}\zeta_2^2+32\zeta_2\zeta_3+240\zeta_5\right)\nonumber\\
&+& C_F C_A^2 \left({1045955 \over 729} + {34732 \over 81} \zeta_2 -\frac{34928}{27} 
\zeta_3-\frac{188}{3}\zeta_2^2+\frac{176}{3}\zeta_2\zeta_3+272\zeta_5\right)\nonumber\\
&+& C_F^2 n_f  \left( - {3826 \over 27} + {296 \over 3} 
\zeta_2-\frac{1232}{9}\zeta_3+\frac{208}{15}\zeta_2^2\right)+C_F n_f^2  \left( {19676 \over 729} + {304 \over 27} 
\zeta_2+\frac{32}{27}\zeta_3\right)\nonumber
\\
&+&C_FC_A n_f  \left( - {309838 \over 729} - {11728 \over 81} 
\zeta_2+\frac{1448}{9}\zeta_3+\frac{88}{15}\zeta_2^2\right)\label{G3}\ ,
\end{eqnarray}
so that we indeed get
\begin{equation}
G_3=H_3-B_3\ ,
\end{equation}
 checking eq.(\ref{check})  for $i=3$. We again note the simple relations between the $C_F^3$ terms in
eqs.(\ref{B3-delta}), (\ref{h3ter}), (\ref{B3-stan}) and (\ref{G3}).
\\

\underline {General structure of $H(a_s)$ and $B(a_s)$}:
\\

\noindent We observe that
\begin{eqnarray} 
C_{DIS}(a_s)+A(a_s)\gamma_E-\beta(a_s)
\frac{\partial}{\partial
a_s}\ln g_0^{DIS}(1,a_s)&\equiv& \beta(a_s)
\frac{\partial}{\partial a_s}\Delta_{DIS}(a_s)\label{Delta}\\
=-\beta_0\Delta_1^{DIS}
a_s^2&-&(\beta_1\Delta_1^{DIS}+2\beta_0\Delta_2^{DIS})a_s^3+...\ ,\nonumber
\end{eqnarray} 
where
$\Delta_{DIS}(a_s)=\Delta_1^{DIS}a_s+\Delta_2^{DIS} a_s^2+...$, and the beta function
 factorizes, in the sense
that the
$\Delta_i^{DIS}$'s are   group theory factors polynomials:
\begin{eqnarray}\label{Delta-i}
\Delta_1^{DIS}&=&C_F(9+4\zeta_2)\nonumber\\
\Delta_2^{DIS}&=&C_F\left[\beta_0\left(
\frac{457}{24}
+19\zeta_2+2\zeta_3\right)\label{Delta-i}\right.\\
&+&\left.C_F\left(-\frac{7}{8}-\frac{75}{2}\zeta_2+66\zeta_3+\frac{6}{5}\zeta_2^2\right)
+ C_A\left(\frac{73}{12}
+14\zeta_2-54\zeta_3-\frac{71}{5}\zeta_2^2
\right)\right]\ .\nonumber
\end{eqnarray}
We thus obtain the general structure 
\begin{equation}
H(a_s)=B_{\delta}(a_s)-\beta(a_s)
\frac{\partial}{\partial a_s}\Delta_{DIS}(a_s)\label{H-stan-bis}\ .
\end{equation}
From
eqs.(\ref{B-stan}) and (\ref{H-stan-bis})   we further obtain the  general expression for
$B$:
\begin{equation}B(a_s)=B_{\delta}(a_s)-G(1,a_s)-\beta(a_s)
\frac{\partial}{\partial a_s}\Delta_{DIS}(a_s)\label{B-stan-bis}\ ,
\end{equation}
which actually  allows to compute  $B_i$ (and in particular  $B_3$)  given the universal virtual
quantities
$B_i^\delta$,
$G_i$, and lower order coefficients with $j<i$ contained in $\Delta_{DIS}(a_s)$.
We also note that
 the  combination 
$B_4-B_4^{\delta}+G_{4}$ among $i=4$ (not yet computed) coefficients can be determined in terms of known
$i\leq 3$ coefficients. Eq.(\ref{B-stan-bis}) is a new result of the present approach.

\section{Threshold resummation of the physical anomalous dimension (DY case)}
In the DY case, the analogue of the resummation formula eq.(\ref{resum}) for the short distance coefficient
function is 
\begin{equation}
\sigma_{DY}(Q^2,N,\mu^2)\sim g_{DY}(Q^2,\mu^2)\ \exp[E_{DY}(Q^2,N,\mu^2)]
\label{resum-DY}\ ,
\end{equation}
with
\begin{equation}
E_{DY}(Q^2,N,\mu^2)=\int_0^1\!\!dz\
2\frac{z^{N-1}-1}{1-z}\left[\int_{\mu^2}^{(1-z)^2Q^2}\!\!\frac{dk^2}{k^2}
A\left(a_s(k^2)\right)+\frac{1}{2}D\left(a_s((1-z)^2Q^2)\right)\right]\label{exponent-DY},
\end{equation}
where
\begin{equation}\label{D-stan}
D(a_s)=
\sum_{i=1}^\infty D_i a_s^{i}\ 
\end{equation}
is the standard Sudakov  anomalous dimension which controls large angle soft gluon emission in the DY process,
whereas
\begin{equation}\label{g-stan-DY}
g_{DY}(Q^2,\mu^2)=1+\sum_{i=1}^\infty  g_i^{DY}\left(\frac{Q^2}{\mu^2}\right)
a_s^{i}(\mu^2)
\end{equation}
collects the constant terms not included in $E_{DY}$. Taking the logarithmic derivative of 
eq.(\ref{resum-DY}) one gets at large-$N$
\begin{equation}\label{scaling-violation-DY}
{d\ln \sigma_{DY}(Q^2,N,\mu^2)\over d\ln Q^2}\sim\int_{0}^1
dz\ 2{z^{N-1}-1
 \over
1-z} {\cal S}[(1-z)^2Q^2]+K\left(a_s(Q^2)\right)\ ,
\end{equation}
where the ``Sudakov effective charge''
\begin{equation} 
 {\cal S}(k^2)=
A\left(a_s(k^2)\right)+\frac{1}{2}{dD\left(a_s(k^2)\right)\over d\ln
k^2}\label{standard-S-coupling-DY}
\end{equation} 
and the ``leftover'' constant terms function (not to be confused with the form factor related $K$ counterterm in eq.(\ref{evolution}))
\begin{equation}
 K\left(a_s(Q^2)\right)={d\ln g_{DY}(Q^2,\mu^2)\over d\ln
Q^2}\label{H-stan-DY}
\end{equation} 
are renormalization group invariant quantities, and ${\cal S}$ refers to the ``soft'' scale $(1-z)^2Q^2$
 in eq.(\ref{scaling-violation-DY}).
Changing variables to
$k^2=(1-z)^2Q^2$, eq.(\ref{scaling-violation-DY}) becomes
\begin{equation}\label{ren-integral-stan-DY}
{d\ln \sigma_{DY}(Q^2,N,\mu^2)\over d\ln Q^2}\sim
\int_{0}^{Q^2}{dk^2\over k^2} F_{DY}\left(\frac{k}{Q},N\right)
{\cal S}(k^2)+K\left(a_s(Q^2)\right)\ ,
\end{equation}
with 
\begin{equation}
F_{DY}\left(\frac{k}{Q},N\right)=\left(1-\frac{k}{Q}\right)^{N-1}-1
\label{eq:F-stan-DY}\ .
\end{equation}
It was further shown in \cite{Grunberg:2006hg} that eq.(\ref{ren-integral-stan-DY}) is equivalent, up to 
corrections which vanish for $N\rightarrow\infty$, to
\begin{equation}\label{ren-integral-stan1-DY}
{d\ln \sigma_{DY}(Q^2,N,\mu^2)\over d\ln Q^2}\sim
\int_{0}^{Q^2}{dk^2\over k^2} G_{DY}\left(\frac{N k}{Q}\right)
  {\cal S}(k^2)+K\left(a_s(Q^2)\right)\ ,
\end{equation}
with
\begin{equation}
G_{DY}\left(\frac{N k}{Q}\right)=\exp\left(-\frac{N k}{Q}\right)-1\label{G-stan-DY}\ ,
\end{equation}
where $G_{DY}(N k/Q)$ is obtained by taking the $N\rightarrow \infty$ limit of $F_{DY}(k/Q,N)$
with $N k/Q$ fixed.
The analogues of the large-$N$ relations eqs.(\ref{ren-int-as}), (\ref{ren-int-as-bis}) and
(\ref{d-scale-viol})  are 
\begin{eqnarray}
{d\ln \sigma_{DY}(Q^2,N,\mu^2)\over d\ln Q^2}\sim\int_{0}^{\infty}{dk^2\over k^2}
G_{DY}\left(\frac{N k}{Q}\right) {\cal S}(k^2)\nonumber\\
+\left[ K\left(a_s(Q^2)\right)+\int_{Q^2}^{\infty}{dk^2\over k^2}  {\cal S}(k^2)
\right]\label{eq:ren-int-as-DY}\ ,
\end{eqnarray}
\begin{eqnarray}
{d\ln \sigma_{DY}(Q^2,N,\mu^2)\over d\ln Q^2}\sim\int_{0}^{\infty}{dk^2\over k^2}
\Big[G_{DY}\left(\frac{N k}{Q}\right)+1\Big] {\cal S}(k^2)\nonumber\\
+\left[ K\left(a_s(Q^2)\right)-\int_{0}^{Q^2}{dk^2\over k^2}  {\cal S}(k^2)
\right]\label{eq:ren-int-as-DY-bis}
\end{eqnarray}
and
\begin{eqnarray}
{d^2\ln \sigma_{DY}(Q^2,N,\mu^2)\over (d\ln Q^2)^2}\sim\int_{0}^{\infty}{dk^2\over k^2}
\dot{G}_{DY}\left(\frac{N k}{Q}\right)  {\cal S}(k^2)\nonumber\\
+\Big[{dK\over d\ln Q^2}-  {\cal S}(Q^2)\Big]\label{eq:d-scale-viol-DY}\ ,
\end{eqnarray}
where $\dot{G}_{DY}\equiv- dG_{DY}/ d\ln k^2$, whereas the analogue of eq.(\ref{conjecture}) is
\begin{equation}
{d^2\ln \vert{\cal F}(-Q^2)\vert^2
\over (d\ln Q^2)^2}={dK\over d\ln Q^2}- 
{\cal S}(Q^2)\label{form-factor-DY}\ ,
\end{equation}
 where ${\cal F}(-Q^2)$ is the time-like quark
form factor.

To compute the left-hand side of eq.(\ref{form-factor-DY}), it is convenient to write
\begin{equation}
\ln \vert{\cal F}(-Q^2)\vert^2=\ln \left({\cal F}(Q^2)\right)^2+{\cal
R}\left(a_s(Q^2)\right)\ ,
\end{equation} 
with
\begin{equation}
{\cal R}\left(a_s(Q^2)\right)\equiv \ln\Big\vert\frac{{\cal F}(-Q^2)}{{\cal F}(Q^2)}
\Big\vert^2\label{R}\ ,
\end{equation}
and we thus get, using eqs.(\ref{form-factor-DY}) and (\ref{left})
\begin{equation}\label{A-dD}
{\cal S}(Q^2)=A\left(a_s(Q^2)\right)+\beta\left(a_s(Q^2)\right)\frac{\partial}{\partial
a_s}\left[-G\left(1,a_s(Q^2)\right)-\beta\left(a_s(Q^2)\right)\frac{\partial{\cal
R}}{\partial a_s}\right.
+K\left(a_s(Q^2)\right)\Big]
\ ,
\end{equation}
which implies, comparing with eq.(\ref{standard-S-coupling-DY})
\begin{equation}\label{D}
\frac{1}{2}D(a_s)=-G(1,a_s)
-\beta(a_s)\frac{\partial{\cal
R}}{\partial a_s}
+K(a_s)\ .
\end{equation}
We thus have to check that 
\begin{equation}\label{D-stan-1}
G(1,a_s)+\beta(a_s)\frac{\partial{\cal
R}}{\partial a_s}=K(a_s)-
\frac{1}{2}D(a_s)\ .
\end{equation}
Now $K(a_s)$ can be computed from the analogues of eqs.(\ref{H}) and (\ref{dg0bis})  
which yield:
\begin{equation}\label{H-DY}
K(a_s)+ C_{DY}(a_s)=
2[B_{\delta}(a_s)-A(a_s)\gamma_E]+\beta(a_s)\frac{\partial}{\partial
a_s}\ln g_0^{DY}(1,a_s)\ ,
\end{equation}
where $C_{DY}$ collects the large-$N$ constant terms included in the Sudakov integrals in
eq.(\ref{ren-integral-stan-DY}) or (\ref{ren-integral-stan1-DY}), and the factor of two on the right-hand side arises
because we have two incoming partons. Similarly to eq.(\ref{H-bis})  we thus have
\begin{equation}\label{H-DY-bis}
K(a_s)=
2
B_{\delta}(a_s)-[
C_{DY}(a_s)+2
A(a_s)\gamma_E]+\beta (a_s)\frac{\partial}{\partial a_s}\ln
g_0^{DY}(1,a_s)\ .
\end{equation}
Moreover the analogue of
eq.(\ref{gamma0}) is:
\begin{eqnarray}
C_{DY}(a_s)&=&2 {\cal S}_1 c_0
a_s+(-4\beta_0 {\cal S}_1 c_1+2 {\cal S}_2 c_0)a_s^2\nonumber\\
&+&[8\beta_0^2 {\cal S}_1 c_2-4(\beta_1 {\cal S}_1+2\beta_0 {\cal S}_2)c_1
+2 {\cal S}_3 c_0]a_s^3+...\label{gamma0-DY}\ ,
\end{eqnarray}
where  ${\cal S}_i$ are defined by
\begin{eqnarray}
{\cal S}(k^2)&=&{\cal S}_1
a_s(Q^2)+\left(-\beta_0 {\cal S}_1\ln\left(\frac{k^2}{Q^2}\right)+{\cal S}_2\right)a_s^2(Q^2)
\label{As-DY-logs}\\
&+&\left(\beta_0^2 {\cal S}_1\ln^2\left(\frac{k^2}{Q^2}\right)-\left(\beta_1 {\cal S}_1+2\beta_0
{\cal S}_2\right)\ln\left(\frac{k^2}{Q^2}\right) +{\cal S}_3\right)a_s^3(Q^2)+...\ .\nonumber
\end{eqnarray}
Eq.(\ref{gamma0-DY}) is easily obtained from eq.(\ref{gamma0}) once one notices that
\begin{equation}J_p(N)\equiv
\int_{0}^{Q^2}{dk^2\over k^2}
\Big[\exp\left(-\frac{N k}{Q}\right)-1\Big]\ln^p\left(\frac{k^2}{Q^2}\right)=2^{p+1}I_p(N)\label{I-J}\ .
\end{equation}
Furthermore eq.(\ref{standard-S-coupling-DY}) gives
\begin{equation}
{\cal S}_1= A_1\ ,\label{A1-DY-stan}
\end{equation}
\begin{equation}
{\cal S}_2= A_2-\beta_0\frac{1}{2} D_1\label{A2-DY-stan}
\end{equation}
and
\begin{equation}
{\cal S}_3= A_3-\beta_1\frac{1}{2}D_1-2\beta_0
\frac{1}{2}D_2\label{A3-DY-stan}\ .
\end{equation}
From eq.(\ref{gamma0-DY}) one can then  infer the general structure (similar to eq.(\ref{Delta}))
\begin{eqnarray} 
C_{DY}(a_s)+2 A(a_s)\gamma_E-\beta (a_s)\frac{\partial}{\partial a_s}\ln
g_0^{DY}(1,a_s)&\equiv& \beta(a_s)
\frac{\partial}{\partial a_s}\Delta_{DY}(a_s)\label{Delta-DY}\\
=-\beta_0\Delta_1^{DY}
a_s^2&-&(\beta_1\Delta_1^{DY}+2\beta_0\Delta_2^{DY})a_s^3+...\ ,\nonumber
\end{eqnarray} 
where
$\Delta_{DY}(a_s)=\Delta_1^{DY} a_s+\Delta_2^{DY} a_s^2+...$, and the $\Delta_i^{DY}$'s
are group theory factors polynomials,
which  yields the general structure of $K(a_s)$
\begin{equation}\label{H-DY-stan-bis}
K(a_s)=
2
B_{\delta}(a_s)-\beta(a_s)
\frac{\partial}{\partial a_s}\Delta_{DY}(a_s)\ .
\end{equation}
One finds:

\noindent i)
\begin{equation}\label{Delta-i-DY-1}\Delta_1^{DY}=C_F(16-8\zeta_2)\ ,\end{equation}
where we used \cite{Moch:2005ky}
\begin{equation}\label{D1-stan}
D_1=0
\end{equation}
and 
\begin{equation}
g_{01}^{DY}=C_F\left(-16+8\gamma_E^2+16 \zeta_2\right)\ .  \label{g01-DY}
\end{equation}

\noindent ii)
\begin{eqnarray}
\label{Delta-i-DY-2}\Delta_2^{DY}&=&C_F\left[\beta_0\left(
\frac{127}{4}
-\frac{56}{3}\zeta_2+12\zeta_3\right)\label{h3ter-DY}\right.\\
&+&\left.C_F\left(\frac{1}{4}-58\zeta_2+60\zeta_3+\frac{88}{5}\zeta_2^2\right)
+ C_A\left(\frac{23}{2}
+\frac{8}{3}\zeta_2
-72\zeta_3+\frac{12}{5}\zeta_2^2\right)\right]\ ,\nonumber
\end{eqnarray}
where we used \cite{Moch:2005ky}
\begin{equation}\label{D2}
D_2=C_FC_A\left(-{1616\over27}+{176\over3}\zeta_2+56\zeta_3
\right)+C_Fn_f\left({224\over27}-{32\over3}\zeta_2\right)
\end{equation}
and
\begin{eqnarray}
g_{02}^{DY}&=&C_F^2\left(\frac{511}{4}-128\gamma_E^2+32\gamma_E^4
-198\zeta_2+128\gamma_E^2\zeta_2-60\zeta_3+\frac{552}{5}\zeta_2^2\right)
\nonumber\\
&+&C_FC_A\left(-\frac{1535}{12}+\frac{1616}{27}\gamma_E+\frac{536}{9}\gamma_E^2+\frac{176}{9}\gamma_E^3
+\frac{376}{3}\zeta_2\nonumber\right.\\
&&\ \ \ \ \ \ \ \ \ \ \ \left.-16\gamma_E^2\zeta_2+\frac{604}{9}\zeta_3-56\gamma_E\zeta_3
-\frac{92}{5}\zeta_2^2\right)\nonumber
\\
&+&C_Fn_f\left(\frac{127}{6}-\frac{224}{27}\gamma_E-\frac{80}{9}\gamma_E^2-\frac{32}{9}\gamma_E^3
-\frac{64}{3}\zeta_2+\frac{8}{9}\zeta_3\right)\label{g02-DY}\ .
\end{eqnarray}

\noindent Eq.(\ref{H-DY-stan-bis})  thus yields the following results in low orders:

\noindent For $i=1$
\begin{equation}
K_1=2 B_1^{\delta}\label{h1-DY}\ .
\end{equation}

\noindent For $i=2$
\begin{equation}
K_2=2 B_2^{\delta}+C_F\beta_0(16-8\zeta_2)\nonumber\label{h2-DY-bis}\ .
\end{equation}

\noindent For $i=3$
\begin{eqnarray}
K_3&=&2 B_3^{\delta}+C_F\beta_1\left(16-8\zeta_2\right)+C_F\beta_0^2\left(
\frac{127}{2}
-\frac{112}{3}\zeta_2+24\zeta_3\right)\label{h3ter-DY}\\
&+&C_F\beta_0\left[C_F\left(\frac{1}{2}-116\zeta_2+120\zeta_3+\frac{176}{5}\zeta_2^2\right)
+ C_A\left(23
+\frac{16}{3}\zeta_2
-144\zeta_3+\frac{24}{5}\zeta_2^2\right)\right]\nonumber\ .
\end{eqnarray}

We can now check eq.(\ref{D-stan-1}) for $i\leq 3$, and thus prove eq.(\ref{form-factor-DY}) to order $a_s^4$. We
first note that from \cite{Magnea:1990zb,Moch:2005id} one gets
\begin{eqnarray}
{\cal R}(a_s)&=&3\zeta_2 A_1 a_s+3\zeta_2 (\beta_0 G_1+A_2) a_s^2+...\nonumber\\
&\equiv& r_1 a_s+r_2 a_s^2+...\label{R-bis}\ ,
\end{eqnarray}
which yields
\begin{eqnarray}
r_1&=&12 C_F \zeta_2\nonumber\\
r_2&=&12 C_F \zeta_2\left[\left(\frac{233}{18}-2\zeta_2\right)C_A-\frac{19}{9}n_f\right]\label{r-i}\ .
\end{eqnarray}
Since $D_1=0$,
 the order  $a_s$ contribution to the right-hand side of eq.(\ref{D-stan-1}) reduces to $K_1$, whereas
 ${\cal R}$ does not contribute at this order to the left-hand side, which reduces to $G_1$. Thus one has to
check that
\begin{equation}
G_1=K_1\label{H-1-DY}\ ,
\end{equation}
which is indeed satisfied (see eq.(\ref{h1-DY}) and the relevant expressions in section 2).

\noindent Next we get from eq.(\ref{h2-DY-bis})
\begin{equation}
K_2=C_F
C_A\left({193\over3}-24\zeta_3\right)-C_F n_f\frac{34}{3}+
C_F^2\left(3
 - 24\zeta_2+48 \zeta_3 \right)\label{H-2-DY}\ ,
\end{equation}
whereas the order $a_s^2$ contribution to the left-hand side of eq.(\ref{D-stan-1}) is
$G_2-\beta_0 r_1$.
So we should check whether
\begin{equation}
G_2=K_2-\frac{1}{2}D_2+\beta_0 r_1\label{G2-check-DY}\ ,
\end{equation}
which is also satisfied.

\noindent Finally we have from eq.(\ref{h3ter-DY})
\begin{eqnarray}\nonumber
K_3&=&C_F^3\left(29+36\zeta_2+136\zeta_3+\frac{576}{5}\zeta_2^2-64\zeta_2\zeta_3-480\zeta_5
\right)\\
&+&C_F^2C_A \left(\frac{232}{3}-\frac{2096}{3}\zeta_2+\frac{3008}{3}\zeta_3-\frac{8}{3}\zeta_2^2+32\zeta_2\zeta_3
+240\zeta_5\right)\nonumber\\
&+&C_F
 C_A^2
\left(\frac{3082}{3}-240\zeta_2-\frac{4952}{9}\zeta_3+\frac{68}{5}\zeta_2^2+80\zeta_5\right)\nonumber\\
&+&C_F^2 n_f\left(-\frac{235}{3}+\frac{320}{3}\zeta_2-\frac{512}{3}\zeta_3+\frac{112}{15}\zeta_2^2\right)
+ C_F n_f^2\left(\frac{220}{9}-\frac{32}{3}\zeta_2+\frac{64}{9}\zeta_3\right)\nonumber\\
&+&C_F C_A  n_f\left(-\frac{3052}{9}+\frac{320}{3}\zeta_2+\frac{208}{9}\zeta_3-\frac{8}{5}\zeta_2^2\right)
\label{h3-DY-ter}
\end{eqnarray}
and \cite{Moch:2005ky,Laenen:2005uz}
\begin{eqnarray}\nonumber
D_3&=&C_FC_A^2 \left(-\frac{594058}{729}+\frac{98224}{81}\zeta_2+\frac{40144}{27}\zeta_3-{2992\over15}
\zeta_2^2 -\frac{352}{3}\zeta_2\zeta_3-384\zeta_5\right)
\nonumber\\
&+&C_F^2n_f \left({3422 \over 27}-32 \zeta_2 -\frac{608}{9}\zeta_3- {64 \over 5}\zeta_2^2\right)+C_Fn_f^2
\left(-{3712\over729}+{640\over27}
\zeta_2 +{320\over27} \zeta_3\right)\nonumber\\ 
&+&C_FC_An_f \left({125252\over729}-{29392\over81} \zeta_2 -{2480\over9}
\zeta_3+\frac{736}{15}\zeta_2^2\right)\label{D3-stan}\ .
\end{eqnarray}
Now the order $a_s^3$ contribution to the left-hand side of eq.(\ref{D-stan-1}) is
$G_3-(\beta_1 r_1+2\beta_0 r_2)$.
So we should check whether
\begin{equation}
G_3=K_3-\frac{1}{2}D_3+\beta_1 r_1+2\beta_0 r_2\label{G3-check-DY}\ ,
\end{equation}
which is again satisfied.
\\

\underline {General structure of $D(a_s)$}:
\\

\noindent From
eqs.(\ref{D-stan-1}) and (\ref{H-DY-stan-bis})    we further obtain the following general expression for
$D$ (the analogue of eq.(\ref{B-stan-bis})):
\begin{equation}
\frac{1}{2}D(a_s)=2 B_{\delta}(a_s)-G(1,a_s)-\beta(a_s)\frac{\partial{\cal
R}}{\partial a_s}-\beta(a_s)
\frac{\partial}{\partial a_s}\Delta_{DY}(a_s)\label{D-stan-bis}\ ,
\end{equation}
which allows to compute  $D_i$ given the universal virtual quantities $B_i^\delta$, $G_i$, and
lower order $j\leq i$ coefficients. Eq.(\ref{D-stan-bis}) is actually closely related to results given in
\cite{Moch:2005ky} and   \cite{Laenen:2005uz}. To make contact with the ``universal''
quantities
$f_q(a_s)$ appearing in \cite{Moch:2005ky} and defined in \cite{Ravindran:2004mb,Moch:2005tm}, we can put
\begin{equation}
G(1,a_s)\equiv {\tilde G}(a_s)+\Delta G(a_s)\label{G-tilde}\ ,
\end{equation}
with ${\tilde G}$ as defined in \cite{Laenen:2005uz,Eynck:2003fn}, such that
\begin{equation}
f_q(a_s)\equiv {\tilde G}(a_s)-2 B_{\delta}(a_s)\label{f}\ .
\end{equation}
We  note that $\Delta G(a_s)$ is also proportional to the beta function, i.e. has the structure
\begin{equation}
\Delta G(a_s)=\beta(a_s)\frac{\partial\kappa}{\partial a_s}\label{Delta-G}\ ,
\end{equation}
where $\kappa(a_s)$ is  a power series with polynomial dependence on the group theory factors. Then
eq.(\ref{D-stan-bis}) becomes
\begin{equation}
\frac{1}{2}D(a_s)=-f_q(a_s)-\beta(a_s)\frac{\partial}{\partial
a_s}\left[{\cal R}(a_s)+\kappa(a_s)+
\Delta_{DY}(a_s)\right]\label{D-stan-ter}\ ,
\end{equation}
which should  be equivalent to eq.(4.4) in \cite{Laenen:2005uz},
and reproduces eq.(36) in \cite{Moch:2005ky}.

As a last comment, we note that subtracting eq.(\ref{B-stan-bis}) from eq.(\ref{D-stan-bis}) we obtain
\begin{equation}
\frac{1}{2}D(a_s)-B(a_s)-B_{\delta}(a_s)=-\beta(a_s)\frac{\partial}{\partial
a_s}\left[{\cal R}(a_s)+
\Delta_{DY}(a_s)-\Delta_{DIS}(a_s)\right]\label{D-B}\ ,
\end{equation}
which yields the relations \cite{Moch:2005ba}
\begin{eqnarray}
\frac{1}{2}D_1-B_1-B_1^{\delta}&=&0\nonumber,\\
\frac{1}{2}D_2-B_2-B_2^{\delta}&=&7 C_F \beta_0\label{D-B-1}\ ,
\end{eqnarray}
as well as the new relation
\begin{eqnarray}\frac{1}{2}D_3-B_3-B_3^{\delta}&=&7 C_F
\beta_1+\beta_0\left[C_AC_F\left(\frac{65}{6}+\frac{28}{3}\zeta_2-36\zeta_3-\frac{75}{5}\zeta_2^2\right)\right.\nonumber\\
+\beta_0C_F\left(\frac{305}{12}\right.&+&\left.\left.\frac{2}{3}\zeta_2+20\zeta_3\right)+C_F^2\left(\frac{9}{4}-41\zeta_2-12\zeta_3+\frac{164}{5}\zeta_2^2\right)\right]\label{D-B-2}\end{eqnarray}
allowing to compute $D_3$ given $B_3$, $B_3^{\delta}$, and information obtained from $i\leq 2$
coefficients. We also note that the combination $\frac{1}{2}D_4-B_4-B_4^{\delta}$ can be determined
in terms of known $i\leq 3$ coefficients.

\section{The variety of resummation procedures}

\subsection{DIS case}
It was   observed in \cite{Grunberg:2006hg,Grunberg:2006jx}
that the separation
between the constant terms contained in the Sudakov integrals on the right-hand side of
eq.(\ref{ren-integral-stan}) or (\ref{ren-integral-stan2})
 and the ``leftover'' constant terms contained in $H(a_s)$ is arbitrary,
yielding a variety of Sudakov resummation  procedures, different choices leading to a different ``Sudakov
distribution function''
$G_{DIS}^{new}(N k^2/Q^2)$ and ``Sudakov effective coupling''
${\cal J}_{new}(k^2)$, as well as to a different function
$H_{new}(a_s)$, namely we  have the alternative large-$N$ representations (up to terms which vanish order by order
in perturbation theory for
$N\rightarrow\infty$)
\begin{equation}\label{scaling-violation2-new}
{d\ln F_2(Q^2,N)\over d\ln Q^2}\sim S_{DIS}^{new}(Q^2,N)+
H_{new}\left(a_s(Q^2)\right)\ ,
\end{equation} 
with
\begin{equation}\label{ren-integral-new} 
S_{DIS}^{new}(Q^2,N)=\int_{0}^{Q^2}{dk^2\over k^2}
G_{DIS}^{new}\left(\frac{N k^2}{Q^2}\right)
 {\cal J}_{new}(k^2)\ .
\end{equation}
As in eq.(\ref{ren-integral1}), we can  extend to infinity the upper limit of integration to obtain
\begin{eqnarray}\label{ren-integral1-new} 
S_{DIS}^{new}(Q^2,N)&\sim&\int_{0}^{\infty}{dk^2\over k^2} G_{DIS}^{new}\left(\frac{N k^2}{Q^2}\right) {\cal J}_{new}(k^2)- G_{DIS}^{new}(\infty)\int_{Q^2}^{\infty}{dk^2\over k^2} {\cal
J}_{new}(k^2)\nonumber\\ &=&
\int_{0}^{\infty}{dk^2\over k^2} G_{DIS}^{new}\left(\frac{N k^2}{Q^2}\right) {\cal J}_{new}(k^2)+\int_{Q^2}^{\infty}{dk^2\over k^2}
{\cal J}_{new}(k^2)\ ,
\end{eqnarray}
 where 
 in the second line, we used that  $G_{DIS}^{new}(\infty)=-1$ for all
resummation procedures, corresponding to the virtual contribution in the Sudakov integral (which determines the
leading logs of
$N$). Thus we get
\begin{equation}
\int_{0}^{\infty}{dk^2\over k^2} G_{DIS}^{new}\left(\frac{N k^2}{Q^2}\right) {\cal J}_{new}(k^2)
+\int_{Q^2}^{\infty}{dk^2\over k^2}
{\cal J}_{new}(k^2)=\sum_{i=1}^\infty \gamma_i^{new}(N)\
a_s^{i}(Q^2)\label{eq:S-series-new}\ ,
\end{equation} 
with ($L\equiv \ln N$)
\begin{eqnarray}
\gamma_1^{new}(N)&=&\gamma_{11}L+\gamma_{10}^{new}\nonumber\\
\gamma_2^{new}(N)&=&\gamma_{22}L^2+\gamma_{21}L+\gamma_{20}^{new}\nonumber\\
\gamma_3^{new}(N)&=&\gamma_{33}L^3+\gamma_{32}L^2+\gamma_{31}L+\gamma_{30}^{new}\label{gammas-new}\\
&etc.&\nonumber\ ,
\end{eqnarray}
where only the {\em non-logarithmic} $\gamma_{i0}^{new}$ terms do depend upon the resummation procedure.

\noindent Alternatively, as in the standard case, one may  remove the virtual contribution from the
Sudakov integral (so that it contains only
 real gluon emission contributions), and merge it together  with the ``leftover'' constant terms,
which yields the equivalent result, in terms of two separately IR divergent (but UV finite) integrals
\begin{equation}\label{ren-integral2-new}
S_{DIS}^{new}(Q^2,N)\sim\int_{0}^{\infty}{dk^2\over k^2} \left[G_{DIS}^{new}\left(\frac{N k^2}{Q^2}\right)+1\right] {\cal J}_{new}(k^2)- \int_{0}^{Q^2}{dk^2\over k^2} {\cal J}_{new}(k^2)\ .
\end{equation}
Using eq.(\ref{ren-integral1-new}) into eq.(\ref{scaling-violation2-new}), we thus end up with the large-$N$
expression
\begin{eqnarray}
{d\ln F_2(Q^2,N)\over d\ln Q^2}&\sim&\int_{0}^{\infty}{dk^2\over k^2} G_{DIS}^{new}\left(\frac{N k^2}{Q^2}\right){\cal J}_{new}(k^2)\nonumber\\ 
&+&\left[H_{new}\left(a_s(Q^2)\right)+\int_{Q^2}^{\infty}{dk^2\over k^2}
{\cal J}_{new}(k^2)\right]\label{ren-int-as-new}\ .
\end{eqnarray} 
If instead one  uses eq.(\ref{ren-integral2-new}) into eq.(\ref{scaling-violation2-new}) one gets the equivalent
form
\begin{eqnarray}
{d\ln F_2(Q^2,N)\over d\ln Q^2}&\sim&\int_{0}^{\infty}{dk^2\over k^2} \left[G_{DIS}^{new}\left(\frac{N k^2}{Q^2}\right)+1\right]{\cal J}_{new}(k^2)\nonumber\\ 
&+&\left[H_{new}\left(a_s(Q^2)\right)-\int_{0}^{Q^2}{dk^2\over k^2}
{\cal J}_{new}(k^2)\right]\label{ren-int-as-bis-new}\ .
\end{eqnarray}
Again we observe \cite{Grunberg:2006gd} that the UV (respectively IR) divergences present in the individual
integrals in eq.(\ref{ren-int-as-new}) (respectively eq.(\ref{ren-int-as-bis-new}))  disappear after taking one
more derivative (which eliminates the virtual contribution inside the Sudakov integral), namely
\begin{eqnarray}
{d^2\ln F_2(Q^2,N)\over (d\ln Q^2)^2}\sim\int_{0}^{\infty}{dk^2\over k^2}
\dot{G}_{DIS}^{new}\left(\frac{N k^2}{Q^2}\right) {\cal J}_{new}(k^2)\nonumber\\
+\left[\frac{dH_{new}}{d\ln Q^2}-{\cal J}_{new}(Q^2)\right]\label{d-scale-viol-new}\ ,
\end{eqnarray}
where $\dot{G}_{DIS}^{new}\equiv- dG_{DIS}^{new}/ d\ln k^2$,  and  the integral in eq.(\ref{d-scale-viol-new}) 
\begin{equation}
S^{'}_{DIS}\left(\frac{Q^2}{N}\right)\equiv\int_{0}^{\infty}{dk^2\over k^2}
\dot{G}_{DIS}^{new}\left(\frac{N k^2}{Q^2}\right) {\cal J}_{new}(k^2)\label{S-dot-exp-new}
\end{equation}
is finite, and consequently  uniquely determined, without need for a ``$new$'' subscript anymore.
The point is that $ S^{'}_{DIS}(Q^2/N)$ being UV (and IR) convergent, all the large-$N$ logarithmic
terms (which are unambiguously fixed) are now determined by the
constant terms contained in the integral, which  cannot be  fixed arbitrarily anymore.
This
 observation implies in turn that  the combination
$dH_{new}/ d\ln Q^2- {\cal J}_{new}(Q^2)$, which represents the ``leftover'' constant terms not included 
in $ S^{'}_{DIS}(Q^2/N)$, is also uniquely fixed.
Consequently
\cite{Grunberg:2006gd}  the conjecture eq.(\ref{conjecture}) has an analogue for {\em all} resummation
procedures, namely
\begin{equation}\label{conjecture-new}
{d^2\ln \left({\cal F}(Q^2)\right)^2 \over
(d\ln Q^2)^2}=\frac{dH_{new}}{d\ln Q^2}- {\cal J}_{new}(Q^2)\ .
\end{equation}
The same unicity statements are actually valid in a
more formal\footnote{A rigorous definition can be given in term of Borel transforms, which are however singular at
 the origin.} sense (since they are UV or IR divergent quantities) for the integrals (which could both be referred
to as
$ S_{DIS}(Q^2/N)$) appearing on the first line of eq.(\ref{ren-int-as-new}) or (\ref{ren-int-as-bis-new}), as well
as for the combination of constant terms appearing on the second line of these equations. In particular, the
second line of eq.(\ref{ren-int-as-bis-new})  can be formally identified, as in the standard procedure (section 2)
to the first logarithmic derivative of the square of the quark form factor, an IR divergent\footnote{Eq.(\ref
{conjecture-0-new}) may still make
 sense at the {\em non-perturbative} level, if one assumes
 that the Sudakov effective charge ${\cal J}_{new}(k^2)$ vanishes for $k^2\rightarrow 0$. This vanishing 
also occurs \cite{Magnea:2000ss,Magnea:2001ge} for $D>4$ in the dimensional regularization framework.}
 quantity:
 \begin{equation}\label{conjecture-0-new}
 {d\ln \left({\cal
F}(Q^2)\right)^2 \over d\ln Q^2}=H_{new}\left(a_s(Q^2)\right)-\int_{0}^{Q^2}{dk^2\over k^2}
{\cal J}_{new}(k^2)\ .
\end{equation}
An application of this relation is given in section 5 to the ``Minkowskian'' resummation formalism.

\noindent Finally, following steps analoguous to those which lead to eq.(\ref{A-dB}), one can  show that 
eq.(\ref{conjecture-new}) is equivalent to
\begin{eqnarray}\label{A-dB-new}
 {\cal J}_{new}(Q^2)&=&A\left(a_s(Q^2)\right)+\beta\left(a_s(Q^2)\right)\frac{\partial}{\partial
a_s}\left[-G\left(1,a_s(Q^2)\right)+H_{new}\left(a_s(Q^2)\right)\right]\nonumber\\
&\equiv& A\left(a_s(Q^2)\right)+\frac{dB_{new}\left(a_s(Q^2)\right)}{d\ln Q^2}
\ ,\end{eqnarray}
which implies the new Sudakov anomalous dimension $B_{new}$ should be given by
\begin{equation}\label{B-new}
B_{new}(a_s)=-G(1,a_s)+H_{new}(a_s)\ .
\end{equation}
Eq.(\ref{A-dB-new}) also shows that the Sudakov effective coupling ${\cal J}_{new}$ differs from the cusp
anomalous dimension by a term proportional to the beta function in {\em all} \footnote{We note
however that the general structures eqs.(\ref{H-stan-bis}) and (\ref{B-stan-bis}) hold usually {\em only} for
the standard procedure, since they rely on the specific relations eqs.(\ref{c0}), (\ref{c1}) and (\ref{c2}).}
 resummation procedures.

\subsection{DY case}
Similar generalizations apply in the DY case. The generalizations of the large-$N$ relations
eqs.(\ref{eq:ren-int-as-DY}), (\ref{eq:ren-int-as-DY-bis})  and (\ref{eq:d-scale-viol-DY})  are 
\begin{eqnarray}
{d\ln \sigma_{DY}(Q^2,N,\mu^2)\over d\ln Q^2}\sim\int_{0}^{\infty}{dk^2\over k^2}
G_{DY}^{new}\left(\frac{N k}{Q}\right) {\cal S}_{new}(k^2)\nonumber\\
+\left[ K_{new}\left(a_s(Q^2)\right)+\int_{Q^2}^{\infty}{dk^2\over k^2}  {\cal S}_{new}(k^2)
\right]\label{eq:ren-int-as-DY-new}\ ,
\end{eqnarray}
\begin{eqnarray}
{d\ln \sigma_{DY}(Q^2,N,\mu^2)\over d\ln Q^2}\sim\int_{0}^{\infty}{dk^2\over k^2}
\Big[G_{DY}^{new}\left(\frac{N k}{Q}\right)+1\Big] {\cal S}_{new}(k^2)\nonumber\\
+\left[ K_{new}\left(a_s(Q^2)\right)-\int_{0}^{Q^2}{dk^2\over k^2}  {\cal S}_{new}(k^2)
\right]\label{eq:ren-int-as-DY-bis-new}
\end{eqnarray}
and
\begin{eqnarray}
{d^2\ln \sigma_{DY}(Q^2,N,\mu^2)\over (d\ln Q^2)^2}\sim\int_{0}^{\infty}{dk^2\over k^2}
\dot{G}_{DY}^{new}\left(\frac{N k}{Q}\right)  {\cal S}_{new}(k^2)\nonumber\\
+\Big[{dK_{new}\over d\ln Q^2}-  {\cal S}_{new}(Q^2)\Big]\label{eq:d-scale-viol-DY-new}\ ,
\end{eqnarray}
where $\dot{G}_{DY}^{new}\equiv- dG_{DY}^{new}/ d\ln k^2$, whereas  the conjecture
eq.(\ref{form-factor-DY}) implies
\begin{equation}
{d^2\ln \vert{\cal F}(-Q^2)\vert^2
\over (d\ln Q^2)^2}={dK_{new}\over d\ln Q^2}- 
{\cal S}_{new}(Q^2)\label{form-factor-DY-new}\ .
\end{equation}
 Eq.(\ref{form-factor-DY-new}) is equivalent to the statement that, for any resummation procedure
\begin{eqnarray}\label{A-dD-new}
{\cal S}_{new}(Q^2)&=&A\left(a_s(Q^2)\right)+\beta\left(a_s(Q^2)\right)\frac{\partial}{\partial
a_s}\left[-G\left(1,a_s(Q^2)\right)-\beta\left(a_s(Q^2)\right)\frac{\partial{\cal
R}}{\partial a_s}\nonumber\right.\\
&+&K_{new}\left(a_s(Q^2)\right)\Big]\nonumber\\
&\equiv&A\left(a_s(Q^2)\right)+\frac{1}{2}\frac{dD_{new}\left(a_s(Q^2)\right)}{d\ln Q^2}\ ,
\end{eqnarray}
with\footnote{Again (footnote 6), the general structure eq.(\ref{D-stan-bis}) is usually valid only for the
standard resummation procedure.}
\begin{equation}\label{D}
\frac{1}{2}D_{new}(a_s)=-G(1,a_s)
-\beta(a_s)\frac{\partial{\cal
R}}{\partial a_s}
+K_{new}(a_s)\ .
\end{equation}

\section{All-order check at large-$n_f$: connection with the dispersive approach}

\subsection{DIS case}
At large-$n_f$ (``large-$\beta_0$'' limit) and finite $N$, the following dispersive representation holds
\cite{Dokshitzer:1995qm,Ball:1995ni}
\begin{equation}
\left.{d\ln F_2(Q^2,N)\over d\ln Q^2}\right\vert_{\textrm{\tiny{large-$\beta_0$}}}=\int_{0}^{\infty}{d\lambda^2\over \lambda^2}
\ddot{{\cal F}}_{DIS}\left({\lambda^2\over Q^2},N\right) A_{Mink}^{V}(\lambda^2)\label{dispersive}\ ,
\end{equation}
 where
\begin{equation}
\frac{1}{4C_F}A_{Mink}^{V}(\lambda^2)={1\over\beta_0}\left[{1\over
2}-{1\over\pi}\arctan\left(\frac{1}{\pi}\ln \left(\frac{\lambda^2}{\Lambda^2_V}\right)\right)\right]\label{eq:A-simple}
\end{equation}
 is the time-like (integrated) discontinuity of the {\em Euclidean} one-loop
coupling (the  ``V-scheme'' coupling) associated to the  dressed gluon propagator
\begin{equation}\frac{1}{4C_F}A_{Eucl}^{V}(k^2)=\frac{1} {\beta_0 \ln
\left(\frac{k^2}{\Lambda^2_V}\right)}\label{one-loop}\ ,
\end{equation} 
($\Lambda_V$ is the V-scheme
scale parameter). Eq.(\ref{dispersive}) represents the ``single dressed gluon'' exchange contribution, i.e. the
infinite sum of diagrams with a single gluon exchange, dressed with an arbitary number of  (renormalized) quark 
loops. Let us now take the large-$N$ limit of eq.(\ref{dispersive}). We shall  use the following two properties of
the Mellin space characteristic function
\begin{equation}
{\cal F}_{DIS}(\epsilon,N)\equiv\int_{0}^1dx\ x^{N-1} \tilde{\cal
F}_{DIS}(\epsilon,x)\label{N-space-char}\ ,
\end{equation}
where $\epsilon\equiv\lambda^2/Q^2$, and $\tilde{\cal F}_{DIS}(\epsilon,x)$ is the momentum space characteristic
function:

\noindent i) For $N\rightarrow\infty$ with $\epsilon_j\equiv N\epsilon$ fixed, we have the scaling property
\cite{Grunberg:2006gd,Grunberg:2006hg,Grunberg:2006jx} (see Appendix B)
\begin{equation}
\ddot{{\cal F}}_{DIS}(\epsilon,N)\sim \ddot{{\cal
G}}_{DIS}(\epsilon_j)\label{G-scaling}\ .
\end{equation}

\noindent ii) For $N\rightarrow\infty$ with $\epsilon$ fixed we have
\begin{equation}
\ddot{{\cal F}}_{DIS}(\epsilon,N)\sim
\ddot{{\cal V}}_s(\epsilon)\label{F-large-N}\ ,
\end{equation} 
where \footnote{Our normalization of ${\cal F}_{DIS}$ and
of
${\cal V}_s$ is half that in \cite{Dokshitzer:1995qm}.}
\cite{Dokshitzer:1995qm}
\begin{equation}
{\cal V}_s(\epsilon)=-\int_0^1 dz {(1-z)^2\over z-\epsilon}\ln{z\over \epsilon}\label{nu}
\end{equation}
is the virtual contribution to the characteristic function. 
Eq.(\ref{F-large-N}) follows from the expression \cite{Dokshitzer:1995qm} for the momentum space  
characteristic function
\begin{equation}
\tilde{\cal F}_{DIS}(\epsilon,x)=\tilde{\cal F}_{DIS}^{(r)}(\epsilon,x)\Theta(1-x-\epsilon
x)+{\cal V}_s(\epsilon)\delta(1-x)\label{char-x-space}\ ,
\end{equation}
(where $\tilde{\cal F}_{DIS}^{(r)}(\epsilon,x)$ is the real contribution) which gives in Mellin space
\begin{equation}
{\cal F}_{DIS}(\epsilon,N)=\int_{0}^{\frac{1}{1+\epsilon}}dx\ x^{N-1} \tilde{\cal
F}_{DIS}^{(r)}(\epsilon,x) +{\cal V}_s(\epsilon)\label{N-space-char-1}\ .
\end{equation} 
To derive the large-$N$ limit of
eq.(\ref{dispersive}), we first take the limit $N\rightarrow\infty$ with $N\epsilon$ fixed inside the dispersive
 integral,  getting
\begin{equation}
\left.{d\ln F_2(Q^2,N)\over d\ln Q^2}\right\vert_{\textrm{\tiny{large-$\beta_0$}}}\sim\int_{0}^{\infty}{d\lambda^2\over
\lambda^2} \ddot{{\cal G}}_{DIS}(N\epsilon) A_{Mink}^{V}(\lambda^2)\label{dispersive-large-N-1}
\ ,
\end{equation}
where the right-hand side is however UV divergent, since $\ddot{{\cal G}}_{DIS}(\infty)=-1$ (corresponding to
 the virtual contribution). To introduce the required UV subtraction, we write eq.(\ref{dispersive}) identically as
\begin{eqnarray}
\left.{d\ln F_2(Q^2,N)\over d\ln Q^2}\right\vert_{\textrm{\tiny{large-$\beta_0$}}}&=&\int_{0}^{\infty}{d\lambda^2\over \lambda^2}
\ddot{{\cal G}}_{DIS}(N\epsilon) A_{Mink}^{V}(\lambda^2)\nonumber\\
&+&\int_{0}^{\infty}{d\lambda^2\over \lambda^2}
\Big[\ddot{{\cal F}}_{DIS}(\epsilon,N)-\ddot{{\cal G}}_{DIS}(N\epsilon)\Big]
A_{Mink}^{V}(\lambda^2)\label{dispersive-bis}\ ,
\end{eqnarray}
and take in a second step the limit  $N\rightarrow\infty$ (with $\epsilon$ fixed!)
 inside the second integral, thus getting\footnote{Eq.(\ref{dispersive-large-N}) is at the basis of the
dispersive approach \cite{Grunberg:2006gd,Grunberg:2006ky} to Sudakov resummation.} 
\begin{equation}
\left.{d\ln F_2(Q^2,N)\over d\ln Q^2}\right\vert_{\textrm{\tiny{large-$\beta_0$}}}\sim\int_{0}^{\infty}{d\lambda^2\over
\lambda^2} \ddot{{\cal G}}_{DIS}(N\epsilon) A_{Mink}^{V}(\lambda^2)
+\int_{0}^{\infty}{d\lambda^2\over \lambda^2}
\Big[\ddot{\cal V}_s(\epsilon)+1\Big] A_{Mink}^{V}(\lambda^2)\label{dispersive-large-N}
\ .
\end{equation}
 The second integral in eq.(\ref{dispersive-large-N}) now appears as an
$N$-independent subtraction term, which regulates the UV divergence of the first integral. It is remarkable, on the
other hand, that both integrals are IR convergent. Indeed, one finds for $\epsilon\rightarrow 0$
\begin{equation}
\ddot{\cal V}_s(\epsilon)+1\sim \epsilon\ln^2\epsilon\label{e-ir}\ ,
\end{equation} 
while, for $\epsilon\rightarrow\infty$, $\ddot{\cal V}_s(\epsilon)={\cal O}(\ln\epsilon/\epsilon)$.
 Comparing with eq.(\ref{ren-int-as-new}) shows
that eq.(\ref{dispersive-large-N}) is nothing but a peculiar case of eq.(\ref{ren-int-as-new}) (at large-$\beta_0$) with ``$new$''=``$Mink$'', provided one makes the identifications
$G_{DIS}^{Mink}(\epsilon_j)\equiv\ddot{{\cal G}}_{DIS}(\epsilon_j)$,
  ${\cal J}_{Mink}(k^2)\vert_{\textrm{\tiny{large-$\beta_0$}}}\equiv A_{Mink}^{V}(k^2)$,
and
identifies the
$N$-independent subtraction term on the second line of eq.(\ref{ren-int-as-new}) as
\begin{equation}
\left.H_{Mink}\left(a_s(Q^2)\right)\right\vert_{\textrm{\tiny{large-$\beta_0$}}}+\int_{Q^2}^{\infty}{dk^2\over
k^2} A_{Mink}^{V}(k^2)\equiv\int_{0}^{\infty}{d\lambda^2\over \lambda^2}
\Big[\ddot{\cal V}_s(\epsilon)+1\Big] A_{Mink}^{V}(\lambda^2)\label{identity-const}\ .
\end{equation}
Eq.(\ref{identity-const}) can be rewritten in terms of UV (and IR)
convergent integrals as
\begin{equation}
\left.H_{Mink}\left(a_s(Q^2)\right)\right\vert_{\textrm{\tiny{large-$\beta_0$}}}=\int_{0}^{Q^2}{d\lambda^2\over \lambda^2}
\Big[\ddot{\cal V}_s(\epsilon)+1\Big] A_{Mink}^{V}(\lambda^2)+\int_{Q^2}^{\infty}{d\lambda^2\over \lambda^2}
\ddot{\cal V}_s(\epsilon) A_{Mink}^{V}(\lambda^2)\label{identity-1}\ ,
\end{equation} 
which gives a dispersive representation of $H_{Mink}$ at large-$\beta_0$ \cite{Grunberg:2006hg}. On the other
hand, both integrals in eq.(\ref{identity-const}) are  UV divergent, but this divergence can be disposed of by
taking one derivative  with respect to $\ln Q^2$, thus getting
\begin{equation}
\left.{dH_{Mink}\left(a_s(Q^2)\right)\over d\ln Q^2}\right\vert_{\textrm{\tiny{large-$\beta_0$}}}-
A_{Mink}^{V}(Q^2)=-\int_{0}^{\infty}{d\lambda^2\over \lambda^2}
\frac{d^3{\cal V}_s(\epsilon)}{(d\ln\epsilon)^3} A_{Mink}^{V}(\lambda^2)\label{d-identity}\ .
\end{equation}
Since, according to our conjecture (eq.(\ref{conjecture-new})), the left-hand side of eq.(\ref{d-identity}) should
be equal to
 ${d^2\ln \left({\cal F}(Q^2)\right)^2 \over (d\ln Q^2)^2}\Big\vert_{\textrm{\tiny{large-$\beta_0$}}}$, we only have to check that
\begin{equation}
\left.{d^2\ln \left({\cal F}(Q^2)\right)^2 \over (d\ln Q^2)^2}\right\vert_{\textrm{\tiny{large-$\beta_0$}}}=-\int_{0}^{\infty}{d\lambda^2\over \lambda^2}
\frac{d^3{\cal V}_s(\epsilon)}{(d\ln\epsilon)^3} A_{Mink}^{V}(\lambda^2)\label{d-identity1}
\end{equation}
is the correct dispersive representation of the second logarithmic derivative of the quark form factor in the large-$\beta_0$ limit. 

\noindent
Paralleling the discussion in sections 2 and 4, we  note that  one can also write the right-hand side of
eq.(\ref{dispersive-large-N}) as the sum of two UV convergent, but IR divergent integrals, by removing the
virtual contribution ($-1$) from the first, $N$-dependent, integral
\begin{equation}
\left.{d\ln F_2(Q^2,N)\over d\ln Q^2}\right\vert_{\textrm{\tiny{large-$\beta_0$}}}\sim\int_{0}^{\infty}{d\lambda^2\over
\lambda^2} \Big[\ddot{{\cal G}}_{DIS}(N\epsilon)+1\Big] A_{Mink}^{V}(\lambda^2)+\int_{0}^{\infty}{d\lambda^2\over
\lambda^2}
\ddot{\cal V}_s(\epsilon) A_{Mink}^{V}(\lambda^2)\label{dispersive-large-N-bis}\ .
\end{equation}
Eq.(\ref{dispersive-large-N-bis}) should be compared to eq.(\ref{ren-int-as-bis-new}), leading to the
identification
\begin{equation}
\left.H_{Mink}\left(a_s(Q^2)\right)\right\vert_{\textrm{\tiny{large-$\beta_0$}}}-\int_{0}^{Q^2}{dk^2\over
k^2} A_{Mink}^{V}(k^2)\equiv\int_{0}^{\infty}{d\lambda^2\over \lambda^2}
\ddot{\cal V}_s(\epsilon) A_{Mink}^{V}(\lambda^2)\label{identity-const-1}\ .
\end{equation}
Comparaison with eq.(\ref{conjecture-0-new}) also  suggests that formally one can also identify the IR divergent 
integral on the right-hand side of eq.(\ref{identity-const-1}) as
\begin{equation}
\left.{d\ln \left({\cal F}(Q^2)\right)^2 \over d\ln Q^2}\right\vert_{\textrm{\tiny{large-$\beta_0$}}}=\int_{0}^{\infty}{d\lambda^2\over \lambda^2}
\ddot{\cal V}_s(\epsilon) A_{Mink}^{V}(\lambda^2)\label{identity-0}\ ,
\end{equation}
a purely virtual contribution, while the first integral on the right-hand side of
eq.(\ref{dispersive-large-N-bis}) can now be interpreted as containing only real gluon emission contributions.

To  check eq.(\ref{identity-0}) (and hence eq.(\ref{d-identity1})), we first observe that the first logarithmic
derivative
$\frac{1}{{\cal F}}\frac{d{\cal F}}{d\ln Q^2}$ of the quark form factor coincides in the large-$\beta_0$ limit with
the ordinary derivative $\frac{d{\cal F}}{d\ln Q^2}$. Indeed, since ${\cal F}=1+{\cal O}(a_s)$, it is clear that
disconnected diagrams coming from the expansion of the denominator $1/{\cal F}$ in the logarithmic derivative are
subdominant at large-$n_f$. Thus the problem  reduces to find the dispersive representation of  $\frac{d{\cal
F}}{d\ln Q^2}$ in the large-$\beta_0$ limit. According to
\cite{Dokshitzer:1995qm,Ball:1995ni}, this amounts to the calculation of the characteristic function $\phi(\lambda^2/Q^2)$ of the quark
form factor, namely of
${\cal F}_1(Q^2)=\phi(\lambda^2/Q^2)\ a_s$, the one loop radiative correction to the on-shell massless quark form factor computed
with a finite gluon mass $\lambda$. This quantity is expected to be UV  divergent, but the divergence should
disappear after taking one derivative, namely $\dot\phi(\lambda^2/Q^2)$ (which is the characteristic function
associated to the derivative of the form factor) should be finite (with $\dot\phi(\lambda^2/Q^2)={\cal
O}(\ln(\lambda^2))$ for $\lambda^2\rightarrow 0$), yielding the dispersive representation
\begin{equation}
\left.\frac{d{\cal F}(Q^2)}{d\ln Q^2}\right\vert_{\textrm{\tiny{large-$\beta_0$}}}=\int_{0}^{\infty}{d\lambda^2\over \lambda^2}
\ddot\phi\left(\frac{\lambda^2}{Q^2}\right)\frac{A_{Mink}^{V}(\lambda^2)}{4C_F}\label{F-dispersive}\ .
\end{equation}
We note that the integral on the right-hand side of eq.(\ref{F-dispersive}) should be UV convergent, but IR divergent,
since we expect $\ddot\phi(\infty)=0$, but
$\ddot\phi(0)={\cal O}(1)$. Comparing eq.(\ref{F-dispersive})  with eq.(\ref{identity-0}) then 
suggests\footnote{The factor $\frac{1}{2}$ in eq.(\ref{result}) arises because the {\em square} of the quark
form factor appears in eq.(\ref{identity-0}).} that
\begin{equation}
\ddot\phi\left(\frac{\lambda^2}{Q^2}\right)=4C_F \frac{1}{2}\ddot{\cal V}_s(\epsilon)=2C_F
\ddot{\cal V}_s(\epsilon)\label{result}\ ,
\end{equation} 
and also that $\dot\phi(\lambda^2/Q^2)=2C_F
\dot{\cal V}_s(\epsilon)$. Moreover
 $2C_F{\cal V}_s(\epsilon)$, which vanishes at $\epsilon=\infty$, should coincide with  the renormalized
version
$\phi_R(\lambda^2/Q^2)$ of
$\phi(\lambda^2/Q^2)$ 
\begin{equation}
2C_F{\cal V}_s(\epsilon)=\phi_R\left(\frac{\lambda^2}{Q^2}\right)\equiv\phi\left(\frac{\lambda^2}{Q^2}\right)-\phi(\infty)\label{result1}\ ,
\end{equation}
with the normalization condition
$\phi_R(\lambda^2/Q^2)=0$ at $Q^2=0$ (i.e. $\phi_R(\infty)=0$). 
These statements are  checked in Appendix C. We also note that ${d\ln \left({\cal F}(Q^2)\right)^2/ d\ln Q^2}$
 has a  status similar to that of an infrared and collinear singular quantity, such as $F_2(Q^2,N)$.

 A partial check of eq.(\ref{d-identity1}) to order $a_s^3$ can also be
performed by comparing the right-hand side expanded to this order with the result of existing order $a_s^3$  
calculations of the quark form factor. This comparaison can be conveniently performed using the Borel transform
technique. Indeed in Borel space eq.(\ref{d-identity1}) becomes (in the $\overline {MS}$ scheme)
\begin{equation}
\left.B\left[{d^2\ln \left({\cal F}(Q^2)\right)^2 \over (d\ln Q^2)^2}\right\vert_{\textrm{\tiny{large-$\beta_0$}}}\right](u)=-4
C_F\exp(5u/3) {\sin\pi u\over\pi u}
\Gamma_{SDG}(u)
\label{B-SDG}\ ,
\end{equation}
where \cite{Grunberg:2006hg}
\begin{equation}
\Gamma_{SDG}(u)=u\int_0^{\infty}\frac{dy}{y}\ddot{\cal V}_s(y)\exp(-u\ln y) 
=\left({\pi u\over\sin\pi
u}\right)^2{1\over (1-u)(1-u/2)}\label{subtraction}\ .
\end{equation}
The left-hand side of eq.(\ref{B-SDG}) is easily obtained to order $u^2$
from eq.(\ref{left}), in the large-$\beta_0$ limit. Our definition of the Borel transform is such that
if $f(a_s)=f_1 a_s+f_2 a_s^2+f_3 a_s^3+...$, then $B[f](u)=f_1 +\frac{f_2}{\beta_0}
u+\frac{1}{2}\frac{f_3}{\beta_0^2} u^2+...$, with $f(a_s)=\frac{1}{\beta_0}\int_0^\infty du\exp(-u/(\beta_0a_s))B[f](u)$.

\subsection{DY case}
We start from the dispersive representation
\begin{equation}
\left.{d\ln \sigma_{DY}(Q^2,N,\mu^2)\over d\ln Q^2}\right\vert_{\textrm{\tiny{large-$\beta_0$}}}=\int_{0}^{\infty}{d\lambda^2\over \lambda^2}
\ddot{{\cal F}}_{DY}\left({\lambda^2\over Q^2},N\right) A_{Mink}^{V}(\lambda^2)\label{dispersive-DY}\ ,
\end{equation}
and take the large-$N$ limit, using the following two scaling properties:

\noindent i) For $N\rightarrow\infty$ with $\epsilon_s^2\equiv N^2\epsilon$ fixed, we have 
\cite{Grunberg:2006gd,Grunberg:2006hg,Grunberg:2006jx} (see Appendix B)
\begin{equation}
\ddot{{\cal F}}_{DY}(\epsilon,N)\sim \ddot{{\cal
G}}_{DY}(\epsilon_s^2)\label{G-scaling-DY}\ .
\end{equation}

\noindent ii) For $N\rightarrow\infty$ with $\epsilon$ fixed we have (see eq.(\ref{N-space-char-1-DY}))
\begin{equation}
\ddot{{\cal F}}_{DY}(\epsilon,N)\sim
\ddot{\cal V}_t(\epsilon)\label{F-large-N-DY}\ ,
\end{equation} 
where  \cite{Dokshitzer:1995qm}
\begin{equation}
{\cal V}_t(\epsilon)=Re\ {\cal V}_s(-\epsilon)=-\int_0^1 dz {(1-z)^2\over z+\epsilon}\ln{z\over
\epsilon}\label{nu-t}\ .
\end{equation}
Proceeding as in section 5.1, we then obtain immediately at large-$N$ the analogues of
eq.(\ref{dispersive-large-N}) and (\ref{dispersive-large-N-bis}): 
\begin{equation}
\left.{d\ln \sigma_{DY}(Q^2,N,\mu^2)\over d\ln Q^2}\right\vert_{\textrm{\tiny{large-$\beta_0$}}}\sim \int_{0}^{\infty}{d\lambda^2\over
\lambda^2} \ddot{{\cal G}}_{DY}(N^2\epsilon) A_{Mink}^{V}(\lambda^2)+\int_{0}^{\infty}{d\lambda^2\over
\lambda^2}
\Big[\ddot{\cal V}_t(\epsilon)+1\Big] A_{Mink}^{V}(\lambda^2)\label{dispersive-large-N-DY}
\ ,
\end{equation}
and
\begin{equation}
\left.{d\ln \sigma_{DY}(Q^2,N,\mu^2)\over d\ln Q^2}\right\vert_{\textrm{\tiny{large-$\beta_0$}}}\sim\int_{0}^{\infty}{d\lambda^2\over
\lambda^2} \Big[\ddot{{\cal G}}_{DY}(N^2\epsilon)+1\Big]
A_{Mink}^{V}(\lambda^2)+\int_{0}^{\infty}{d\lambda^2\over
\lambda^2}
\ddot{\cal V}_t(\epsilon) A_{Mink}^{V}(\lambda^2)\label{dispersive-large-N-bis-DY}\ .
\end{equation}
The result for the time-like form factor is immediately obtained by analytic continuation of eq.(\ref{identity-0})
to the time-like region. Performing this continuation and taking the real part of the result one gets the formal
IR divergent dispersive representation
\begin{equation}
\left.{d\ln \vert{\cal F}(-Q^2)\vert^2 \over d\ln Q^2}\right\vert_{\textrm{\tiny{large-$\beta_0$}}}=\int_{0}^{\infty}{d\lambda^2\over \lambda^2}
\ddot{\cal V}_t(\epsilon) A_{Mink}^{V}(\lambda^2)\label{identity-0-DY}\ ,
\end{equation}
whereas the analogue of eq.(\ref{d-identity1}) is clearly
\begin{equation}
\left.{d^2\ln \vert{\cal F}(Q^2)\vert^2 \over (d\ln Q^2)^2}\right\vert_{\textrm{\tiny{large-$\beta_0$}}}=-\int_{0}^{\infty}{d\lambda^2\over \lambda^2}
\frac{d^3{\cal V}_t(\epsilon)}{(d\ln\epsilon)^3} A_{Mink}^{V}(\lambda^2)\ .\label{d-identity1-DY}
\end{equation}

\section{Conclusion}
In this paper we investigated the structure of $N$-independent contributions in threshold resummation
 for DIS and the DY process.
Our main result  is contained in the conjectured relations eq.(\ref{conjecture}) and (\ref{form-factor-DY}),
 which have been checked to order $\alpha_s^4$, using the  relations eq.(\ref{B-H-bis}) and (\ref{D-stan-1}).
These relations essentially state that, once one corrects for the mismatch due to the presence of a virtual
 contribution
 (required to regulate IR divergencies) in the Sudakov integrals (which ideally should contain only real gluon
emission contributions responsible for the logarithmic terms at large-$N$),  the remaining  constant terms not
included in the  integrals are given by (logarithmic derivatives of) the quark form factor, a purely virtual
contribution. To obtain these relations, two derivatives are necessary (eq.(\ref{d-scale-viol}) and (\ref{eq:d-scale-viol-DY})): the first one gets rid of infrared and collinear divergences, leading to the IR safe
``physical anomalous dimension'' observable; the second one allows to bypass the real-virtual cancellation of IR singularities, the purely real contributions (the Sudakov integrals in  eq.(\ref{d-scale-viol}) and (\ref{eq:d-scale-viol-DY})) and the remainder, purely virtual, form factor related  constant terms being {\em separately} finite.

\noindent The close connection between $N$-independent (``non-logarithmic'')  terms and form factor type
contributions  have been noted for a long time
\cite{Sterman:1986aj,Moch:2005ky,Laenen:2005uz,Eynck:2003fn,Ravindran:2006cg}. We presented a
particularly simple version of this connection,  valid in four dimensions, where the DIS and DY channels are
treated in a symmetrical way. As a by-product, we obtained eq.(\ref{B-stan-bis}) and (\ref{D-stan-bis}), which allow
to compute the ``non-conformal'' parts $B$ and $D$ of the standard ``jet'' and ``soft''  Sudakov
effective charges ${\cal J}$ and ${\cal S}$ in terms  of the virtual contribution $B_{\delta}$ to the
diagonal splitting function, and the quark form factor. While the second of these relations has a content 
equivalent
to similar ones previously given in \cite{Moch:2005ky,Laenen:2005uz,Eynck:2003fn} for the DY process
 (which allow in
particular to compute $D_3$, yielding a result which agrees with the one obtained in
\cite{Moch:2005ky,Laenen:2005uz,Idilbi:2005ni,Ravindran:2005vv}), its counterpart  eq.(\ref{B-stan-bis}) for the
DIS case is new, and allows to compute
$B_3$  with a   method alternative to the one used in \cite{Moch:2005ba}. Moreover, subtracting
 eq.(\ref{B-stan-bis}) 
from (\ref{D-stan-bis}), we obtained the general structure (eq.(\ref {D-B})) laying behind the first relation in 
eq.(4.19) of \cite{Moch:2005ba}, which we extended to  one more order (eq.(\ref{D-B-2})).

\noindent We also performed an all-order check of our conjecture at large-$n_f$, taking the large-$N$ limit of the
dispersive representation of the ``physical anomalous dimensions'' which control the scaling violation for DIS and
DY. As a by-product, we obtained a dispersive representation of the quark form factor.

\noindent A further consequence of eq.(\ref{conjecture}) and (\ref{form-factor-DY}) was pointed out
 in \cite{Grunberg:2006gd}: if the theory is conformal
 in the IR limit, these relations imply {\em universality} of the IR fixed points of the DIS and DY Sudakov
 effective couplings ${\cal J}(k^2)$ and ${\cal S}(k^2)$; in particular, the Banks-Zaks fixed points are the same
 (and independent of the resummation procedure, i.e. the same for ${\cal J}_{new}(k^2)$ and ${\cal S}_{new}(k^2)$
).

\noindent Although we investigated only DIS and DY, we expect a similar approach to be applicable to other inclusive processes with a hard electromagnetic vertex, such as the ones considered in
\cite{Sterman:2006hu}, as well as \cite{Laenen:2005uz} to the case, relevant to Higgs production via gluon fusion,  where the hard vertex is  a gluon form factor.

\vspace{0.5cm}

\noindent\textbf{Acknowledgements}

\vspace{0.3cm}

\noindent G. G. wishes to thank Yu. L. Dokshitzer, E. Gardi,  L. Magnea, G. Marchesini and G. Sterman for instructive discussions. 
S. F. acknowledges financial support from MEC (Spain) grant CYT FPA 2004-04582-C02-01, and the RTN Flavianet MRTN-CT-2006-035482 (EU). He would like to thank \'Ecole Polytechnique for hospitality at CPhT during the year 2005-2006.
\appendix
\section{The large-$N$ constant  terms in the Sudakov integral}

To compute the constant terms $c_p$ in eq.(\ref{I-p}), \textit{i.e} the terms proportional to $N^0$ (or, rather, $N^{-0}$) in the $N\rightarrow\infty$ asymptotic expansions of (\ref{I-p}), it is very convenient to 
introduce the Mellin-Barnes representation
\begin{equation}\label{MBrep}
\exp\left(-\frac{N k^2}{Q^2}\right)-1=\frac{1}{2\pi i}\int_{c-i\infty}^{c+i\infty}ds\left(\frac{Nk^2}{Q^2}\right)^{-s}\Gamma(s)\ ,
\end{equation}
with $c\equiv\Re(s)\in]-1,0[$. The latter interval defines the $s$-complex plane fundamental strip of the Mellin-Barnes representation, crucial object to determine the asymptotic expansion. Indeed, the Mellin transform singularities lying to the right of the fundamental strip encode the $N\rightarrow\infty$ asymptotic expansion of the Mellin-Barnes representation, while the singularities to the left encode the $N\rightarrow0$ asymptotic expansion. It is therefore important to know precisely the fundamental strip in order to take into account only the relevant singularities (for more details, in particular for the determination of the fundamental strip, we refer the reader to \cite{FlaGoDu95,Friot:2005cu}).

Using (\ref{MBrep}) in eq.(\ref{I-p}), we then have 
\begin{eqnarray}
I_0(N)&\equiv&\frac{1}{2\pi i}\int_{c-i\infty}^{c+i\infty}ds\,N^{-s}\tilde{I}_0(s)=\frac{1}{2\pi i}\int_{c-i\infty}^{c+i\infty}ds\left(\frac{N}{Q^2}\right)^{-s}\Gamma(s)
\int_0^{Q^2}dk^2\frac{1}{(k^2)^{1+s}}\nonumber\\
&=&-\frac{1}{2\pi i}\int_{c-i\infty}^{c+i\infty}ds\,N^{-s}\Gamma(s)\frac{1}{s}\ ,\\
I_1(N)&\equiv&\frac{1}{2\pi i}\int_{c-i\infty}^{c+i\infty}ds\,N^{-s}\tilde{I}_1(s)=\frac{1}{2\pi i}\int_{c-i\infty}^{c+i\infty}ds\left(\frac{N}{Q^2}\right)^{-s}\Gamma(s)
\int_0^{Q^2}dk^2\frac{1}{(k^2)^{1+s}}\ln\left(\frac{k^2}{Q^2}\right)\nonumber\\
&=&-\frac{1}{2\pi
i}\int_{c-i\infty}^{c+i\infty}ds\,N^{-s}\Gamma(s)\frac{1}{s^2}
\end{eqnarray}
and 
\begin{eqnarray}
I_2(N)&\equiv&\frac{1}{2\pi i}\int_{c-i\infty}^{c+i\infty}ds\,N^{-s}\tilde{I}_2(s)=\frac{1}{2\pi i}\int_{c-i\infty}^{c+i\infty}ds\left(\frac{N}{Q^2}\right)^{-s}\Gamma(s)
\int_0^{Q^2}dk^2\frac{1}{(k^2)^{1+s}}\ln^2\left(\frac{k^2}{Q^2}\right)\nonumber\\
&=&-\frac{1}{2\pi
i}\int_{c-i\infty}^{c+i\infty}ds\,N^{-s}\Gamma(s)\frac{2}{s^3}\ ,
\end{eqnarray}
where the last step of integration for each integral is true only if $\Re(s)<0$. Since $\Re(s)\in]-1,0[$, this is actually the case. 
For arbitrary $p$ we have
\begin{equation}
I_p(N)\equiv \frac{1}{2\pi i}\int_{c-i\infty}^{c+i\infty}ds\,N^{-s}\tilde{I}_p(s)=-\frac{1}{2\pi
i}\int_{c-i\infty}^{c+i\infty}ds\,N^{-s}\Gamma(s)\frac{p\ !}{s^{p+1}}\ ,
\end{equation}
with the same condition $\Re(s)<0$.

Now, we are interested in the $N\rightarrow\infty$ asymptotic expansion of these integrals
 and, to be precise, in the constant term of the asymptotic expansion, therefore we only have to consider the
corresponding singular element of the $\tilde{I}_p(s)$ (to the right of their
fundamental strip). This is indeed the statement of the \textit{converse mapping theorem}
\cite{FlaGoDu95}, which gives nothing but a simple dictionary between singular elements (or singularities) of the Mellin transforms (here the $\tilde{I}_p(s)$) and terms of the asymptotic expansions of their inverse (here the $I_p(N)$).  
In our case, the singular elements of interest $\tilde{I}_p\vert^{sing.}$ ($p\in\{0,1,2\}$) are the truncated Laurent series of each $\tilde{I}_p(s)$ around $s=0$, because the \textit{converse mapping theorem} relates the residue of a pole located at $s=l$ to the coefficient of the term of power $-l$ in the asymptotic expansion (which is for us $N^{-0}$).

The singular elements at $s=0$ are 
\begin{equation}
\tilde{I}_0\vert^{sing.}_{s\rightarrow0}=-\frac{1}{s^2}+\frac{\gamma_E}{s}\ ,
\end{equation}
\begin{equation}
\tilde{I}_1\vert^{sing.}_{s\rightarrow0}=-\frac{1}{s^3}+\frac{\gamma_E}{s^2}-\frac{6\gamma_E^2+\pi^2}{12s}
\end{equation}
and
\begin{equation}
\tilde{I}_2\vert^{sing.}_{s\rightarrow0}=-\frac{2}{s^4}+\frac{2\gamma_E}{s^3}-\frac{6\gamma_E^2+\pi^2}{6s^2}+\frac{2\gamma_E^3
+\gamma_E\pi^2+4\zeta_3}{6s}\ .
\end{equation}
Since the Mellin transforms $\tilde{I}_p(s)$ fulfil the necessary condition of decrease along vertical lines \cite{FlaGoDu95}, we can apply the \textit{converse mapping theorem}\footnote{The \textit{converse mapping theorem}  says that a pole of multiplicity $m$ located at $s=l$ gives a term proportional to $N^{-l}\ln^{m-1}(N)$ so that we only need to consider the $m=1$ terms in our singular elements.} and we then find the results in eqs.(\ref{c0}), (\ref{c1}) and (\ref{c2}). Notice that with this method any term of the asymptotic expansions can be straightforwardly obtained by computing the corresponding singular element (see Appendix C for an example of a complete asymptotic expansion computation).

\section{Scaling behavior of the characteristic functions in the large-$N$ limit}

\subsection{Deep Inelastic Scattering}

 We start from the expression eq.(\ref{N-space-char-1}) for the
Mellin-space characteristic function $\mathcal{F}_{DIS}(\epsilon,N)$, and derive its $N\rightarrow\infty$ limit, with
$\epsilon_j\equiv N\epsilon=\frac{N\lambda^2}{Q^2}$ fixed. Let us first consider the real contribution
\begin{equation}
{\cal F}_{DIS}^{(r)}(\epsilon,N)=\int_{0}^{\frac{1}{1+\epsilon}}dx\ x^{N-1} \tilde{\cal
F}_{DIS}^{(r)}(\epsilon,x)\label{N-space-char-r}\ .
\end{equation} 
Using the change of variable $t=N(1-x)$, we get
\begin{equation}\label{cdv}
\mathcal{F}_{DIS}^{(r)}(\epsilon,N)=\int_{\frac{\epsilon_j}
{1+\frac{\epsilon_j}{N}}}^{N}dt\left(1-\frac{t}{N}\right)^{N-1}\frac{1}{N}\tilde\mathcal{F}_{DIS}^{(r)}
\left(\frac{\epsilon_j}{N},1-\frac{t}{N}\right)\ .
\end{equation}
Now for $N\rightarrow\infty$, 
$\left(1-\frac{t}{N}\right)^{N-1}\sim \exp(-t)$, whereas, using the expression of $\tilde{\cal
F}_{DIS}^{(r)}(\epsilon,x)$ given in \cite{Dokshitzer:1995qm}, one finds\footnote{We note that after multiplication by $t$,
and reverting to the original variables $x$ and $\epsilon$, the left-hand side of eq.(\ref{F-logs}) coincides with
$(1-x)\tilde{\cal F}_{DIS}^{(r)}(\epsilon,x)$. We thus find  for $x\rightarrow1$
with $\frac{\epsilon}{1-x}$ fixed  the scaling law in momentum space $(1-x)\tilde{\cal
F}_{DIS}^{(r)}(\epsilon,x)\sim 
\ln\frac{1-x}{\epsilon}-\frac{3}{4}+\frac{1}{2}\frac{\epsilon}{1-x}+\frac{1}{4}(\frac{\epsilon}{1-x})^2$,
where the right-hand side depends only on the {\em single} variable $\frac{\epsilon}{1-x}=\frac{\lambda^2}{Q^2(1-x)}$, 
and coincides (taking into account the different normalization) with $(1-x)\mathcal{F}(x,\epsilon)\left.\right\vert_{\textrm{\tiny{log}}}$ in the
notation of
\cite{Gardi:2001di} (see eq.(28) there).}    
\begin{equation}\label{F-logs}
\frac{1}{N}\tilde\mathcal{F}_{DIS}^{(r)}
\left(\frac{\epsilon_j}{N},1-\frac{t}{N}\right)
\sim\frac{1}{t}\left(
\ln\frac{t}{\epsilon_j}-\frac{3}{4}+\frac{1}{2}\frac{\epsilon_j}{t}+\frac{1}{4}\frac{\epsilon_j^2}{t^2}\right)\,,
\end{equation}
(where we  accounted for the different normalization by a factor $1/2$). Thus taking the limit $N\rightarrow\infty$
with $\epsilon_j$ fixed we get
\begin{equation}\label{largeNint}
\mathcal{F}_{DIS}^{(r)}(\epsilon,N)\sim\int_{\epsilon_j}^{+\infty}\frac{dt}{t}\ 
\exp(-t)\left[\ln\frac{t}{\epsilon_j}-
\frac{3}{4}+\frac{1}{2}\frac{\epsilon_j}{t}+\frac{1}{4}\frac{\epsilon_j^2}{t^2}\right]\,,
\end{equation}
where we  set $N=\infty$  in the limits of integration. We are interested in the behavior of the second derivative
 with
respect to $\ln\epsilon_j$. It is straightforward to get from eq.(\ref{largeNint})
\begin{equation}\label{d-largeNint}
\dot{\mathcal{F}}_{DIS}^{(r)}(\epsilon,N)\sim\int_{\epsilon_j}^{+\infty}\frac{dt}{t}\ 
\exp(-t)\left[1-\frac{1}{2}\frac{\epsilon_j}{t}-\frac{1}{2}\frac{\epsilon_j^2}{t^2}\right]
\end{equation}
and
\begin{eqnarray}\label{dd-largeNint}
\ddot{\mathcal{F}}_{DIS}^{(r)}(\epsilon,N)&\sim&\int_{\epsilon_j}^{+\infty}\frac{dt}{t}\ 
\exp(-t)\left[\frac{1}{2}\frac{\epsilon_j}{t}+\frac{\epsilon_j^2}{t^2}\right]\nonumber\\
&=&\frac{\epsilon_j}{2}\Gamma(-1,\epsilon_j)+\epsilon_j^2\Gamma(-2,\epsilon_j)\ ,
\end{eqnarray}
 where we recall that $\dot{\mathcal{F}}\equiv
-\frac{d{\mathcal{F}}}{d\ln\lambda^2}=+\frac{d{\mathcal{F}}}{d\ln Q^2}$.  On the other hand, for
$\epsilon=\epsilon_j/N\rightarrow 0$ the virtual contribution
${\cal V}_s(\epsilon)$ behaves as
\begin{equation}
{\cal V}_s(\epsilon)\sim-\frac{1}{2}\ln^2\epsilon-\frac{3}{2}\ln\epsilon-\frac{\pi^2}{3}-\frac{7}{4}
\label{nu-ir}\ ,
\end{equation}
and thus diverges for $N\rightarrow\infty$ with $\epsilon_j$ fixed, but this divergence is removed after taking two
derivatives, since (see also eq.(\ref{e-ir})) $\ddot{\cal V}_s(0)=-1$. 
We can express the incomplete Gamma functions $\Gamma(-1,\epsilon_j)$ and $\Gamma(-2,\epsilon_j)$ on the right-hand
 side
of eq.(\ref{dd-largeNint}) in terms of
$\Gamma(0,\epsilon_j)$ using integration by parts 
\begin{equation}
\Gamma(-1, x)=\frac{\exp(-x)}{x}-\Gamma(0, x)
\end{equation}
and
\begin{equation}
\Gamma(-2, x)=\frac{1}{2}\frac{\exp(-x)}{x^2}-\frac{1}{2}\frac{\exp(-x)}
{x}+\frac{1}{2}\Gamma(0, x)\ ,
\end{equation}
to obtain
\begin{equation}\label{d-largeNint-1}
\ddot{\mathcal{F}}_{DIS}^{(r)}(\epsilon,N)\sim \exp(-\epsilon_j)-\frac{1}{2}\epsilon_j
\exp(-\epsilon_j)-\frac{1}{2}\epsilon_j\Gamma(0,\epsilon_j)+\frac{1}{2}
\epsilon_j^2\Gamma(0,\epsilon_j)\ .
\end{equation}
Hence, adding the virtual contribution $\ddot{\cal V}_s(0)=-1$, we get
\cite{Grunberg:2006gd,Grunberg:2006hg,Grunberg:2006jx}
\begin{equation}\label{d-largeNint-2}
\ddot{\mathcal{F}}_{DIS}(\epsilon,N)\sim -1+\exp(-\epsilon_j)-\frac{1}{2}\epsilon_j
\exp(-\epsilon_j)-\frac{1}{2}\epsilon_j\Gamma(0,\epsilon_j)+\frac{1}{2}
\epsilon_j^2\Gamma(0,\epsilon_j)\equiv\ddot{\cal G}_{DIS}(\epsilon_j)\ .
\end{equation}

\subsection{Drell-Yan}

Exactly the same reasoning as for DIS can be applied to DY. As before, we use results of \cite{Dokshitzer:1995qm}
 at finite
$N$  to compute the corresponding large-$N$ limit. 
We begin with the analogue of eq.(\ref{N-space-char-1})
\begin{equation}
{\cal F}_{DY}(\epsilon,N)=\int_{0}^{\tau_{max}}dx\ x^{N-1} \tilde{\cal
F}_{DY}^{(r)}(\epsilon,x) +{\cal V}_t(\epsilon)\label{N-space-char-1-DY}\ ,
\end{equation}
where $\tau_{\textrm{max}}=\frac{1}{(1+\sqrt{\epsilon})^2}$, and derive the $N\rightarrow\infty$ limit of ${\cal
F}_{DY}(\epsilon,N)$, with
$\epsilon_s\equiv N\sqrt{\epsilon}=\frac{N\lambda}{Q}$ fixed.
Using again the change of variable $t=N(1-x)$, we get for the real contribution 
\begin{equation}\label{cdv2}
\mathcal{F}_{DY}^{(r)}(\epsilon,N)=\int_{\frac{2\epsilon_s+\frac{\epsilon_s^2}{N}}
{\left(1+\frac{\epsilon_s}{N}\right)^2}}^{N}dt\left(1-\frac{t}{N}\right)^{N-1}\frac{1}{N}\tilde\mathcal{F}_{DY}^{(r)}
\left(\frac{\epsilon_s^2}{N^2},1-\frac{t}{N}\right)\ .
\end{equation}
Starting from the expression of $\tilde{\cal
F}_{DY}^{(r)}(\epsilon,x)$ given in \cite{Dokshitzer:1995qm}, it is easy to show that for $N\rightarrow\infty$ 
 we have\footnote{Again we note that after
multiplication by
$t$, and reverting to the original variables $x$ and $\epsilon$, the left-hand side of eq.(\ref{F-logs-DY}) coincides
with
$(1-x)\tilde{\cal F}_{DY}^{(r)}(\epsilon,x)$. We thus find  for $x\rightarrow1$
with $\frac{\epsilon}{(1-x)^2}$ fixed  the scaling law in momentum space $(1-x)\tilde{\cal
F}_{DY}^{(r)}(\epsilon,x)\sim 4\,\textrm{tanh}^{-1}\sqrt{1-4\frac{\epsilon}{(1-x)^2}}$,
where the right-hand side depends only on the {\em single} variable
 $\frac{\epsilon}{(1-x)^2}=\frac{\lambda^2}{Q^2(1-x)^2}$.} (taking
into account  the different normalization by a factor of $1/2$)   
\begin{equation}\label{F-logs-DY}
\frac{1}{N}\tilde\mathcal{F}_{DY}^{(r)}
\left(\frac{\epsilon_s^2}{N^2},1-\frac{t}{N}\right)
\sim\frac{4}{t}\textrm{tanh}^{-1}\sqrt{1-\frac{4\epsilon_s^2}{t^2}}\ .
\end{equation}
Thus letting $N\rightarrow\infty$
with $\epsilon_s$ fixed we get
\begin{equation}\label{largeNint-DY}
\mathcal{F}_{DY}^{(r)}(\epsilon,N)\sim4\int_{2\epsilon_s}^{+\infty}\frac{dt}{t}\,
\exp(-t)\ \textrm{tanh}^{-1} 
\sqrt{1-\frac{4\epsilon_s^2}{t^2}}\ ,
\end{equation}
where we  set $N=\infty$  in the limits of integration. Taking (minus) the first derivative with respect to
$\ln\epsilon_s^2$ of eq.(\ref{largeNint-DY}) we obtain
\begin{equation}\label{d-largeNint-DY}
\dot{\mathcal{F}}_{DY}^{(r)}(\epsilon,N)\sim2\int_{2\epsilon_s}^{+\infty}\frac{dt}{t}\,
\exp(-t)\  
\frac{1}{\sqrt{1-\frac{4\epsilon_s^2}{t^2}}}=2K_0\left(2\epsilon_s\right)\ ,
\end{equation}
where $K_0$ is the modified Bessel function of the second kind\footnote{A connection with the work in \cite{Laenen:2000ij} was pointed out in \cite{Grunberg:2006gd}.}.
On the other hand, for $\epsilon=\epsilon_s^2/N^2\rightarrow 0$ the virtual contribution ${\cal V}_t(\epsilon)$ behaves as
\begin{equation}
{\cal V}_t(\epsilon)\sim-\frac{1}{2}\ln^2\epsilon-\frac{3}{2}\ln\epsilon+\frac{\pi^2}{6}-\frac{7}{4}
\label{nu-t-ir}\ ,
\end{equation}
and thus diverges for $N\rightarrow\infty$ with $\epsilon_s$ fixed, but this divergence is again removed after taking
 two
derivatives, and we get  $\ddot{\cal V}_t(0)=-1$. Thus we obtain
\begin{equation}
\ddot{\mathcal{F}}_{DY}(\epsilon,N)\sim
-\frac{d}{d\ln\epsilon_s^2}\left[2K_0\left(2\epsilon_s\right)\right] -1=
-\frac{d}{d\ln x}\left[K_0\left(x=2\epsilon_s\right)\right] -1\equiv\ddot{\cal G}_{DY}(\epsilon_s^2)
\label{dd-largeNint-DY}\ ,
\end{equation}
which agrees with the result quoted in
\cite{Grunberg:2006gd,Grunberg:2006hg,Grunberg:2006jx}.

\section{Massless one-loop quark form factor with a finite gluon mass}

Let us detail the calculation of the massless one-loop renormalized quark form factor with a finite gluon mass $\lambda$. 
We present here this calculation in dimensional regularisation $D=4-\epsilon$ and in Feynman gauge
 (the ``Landau gauge'' $k_\mu k_\nu$ term gives no contribution).

The amplitude we are interested in is
\begin{equation}
A=(-ie)(-i)g^2t^at^a\mu^\epsilon
\bar u(p+q)\int\frac{d^{4-\epsilon}k}{(2\pi)^{4-\epsilon}}\ \gamma^\mu\frac{1}{\hat k+\hat p +\hat q 
+i\eta}\gamma_\nu\frac{1}{\hat k+\hat p+i\eta}\gamma_\mu\frac{1}{k^2-\lambda^2+i\eta}\ u(p)
\end{equation}
with $t^at^a=C_F=\frac{4}{3}$.
In the (on-shell) massless quark limit only one form factor enters into the game: 
\begin{equation}
A=(-ie)\mathcal{F}_1(Q^2,\epsilon)\bar u(p+q)\gamma_{\nu}u(p)\ ,
\end{equation}
where $Q^2\equiv -q^2$.

Feynman parametrisation, gamma-algebra and evaluation of the momentum integral lead to (we do not write $i\eta$
 anymore)
\begin{eqnarray}
\mathcal{F}_1(Q^2,\epsilon)&=&(-i)\alpha_s4\pi
C_F\left\{\mu^\epsilon2\frac{i}{(2\sqrt{\pi})^{4-\epsilon}}\frac{\Gamma(3-\frac{\epsilon}{2})
\Gamma(\frac{\epsilon}{2})}{\Gamma(2-\frac{\epsilon}{2})\Gamma(3)}\frac{(\epsilon-2)^2}{4-\epsilon}\right.\nonumber\\
&\times&\!\!\!\int_0^1\!dx\!
\int_0^1\!dy
y[Q^2xy^2(1-x)+\lambda^2(1-y)]^{\frac{-\epsilon}{2}}\nonumber\\
&&\ \ \ \ \ \ \ \ \ \ \ \ \ \ \ \  +\left.\mu^\epsilon2\frac{-i}{(2\sqrt{\pi})^{4-\epsilon}}\frac{\Gamma(1+\frac{\epsilon}{2})}{\Gamma(3)}\right.\nonumber\\
&\times&\!\!\!\left.\int_0^1\!dx\!
\int_0^1\!dy y[Q^2xy^2(1-x)+\lambda^2(1-y)]^{-1-\frac{\epsilon}{2}}Q^2[2(1-y)+(x-1)xy^2(\epsilon-2)]\right\}.\nonumber\\
\label{F1}
\end{eqnarray}
The exact values of these integrals are not straightforwardly computed but one can get their asymptotic expansions
 in the $\lambda\rightarrow 0$ limit. Interestingly enough, we shall see that exact results can be obtained from the
asymptotic expansions. The calculation will be done following the strategy of \cite{FlaGoDu95,Friot:2005cu} which has already been used and detailed in Appendix A.

Let us consider the first parametric integral in the bracket of (\ref{F1}). It reads
$$
I_1\equiv \mu^\epsilon2\frac{i}{(2\sqrt{\pi})^{4-\epsilon}}\frac{\Gamma(3-\frac{\epsilon}{2})
\Gamma(\frac{\epsilon}{2})}{\Gamma(2-\frac{\epsilon}{2})\Gamma(3)}\frac{(\epsilon-2)^2}{4-\epsilon}\int_0^1dx
\int_0^1dy
y[Q^2xy^2(1-x)+\lambda^2(1-y)]^{\frac{-\epsilon}{2}}
$$
\begin{equation}
=\mu^\epsilon2\frac{i}{(2\sqrt{\pi})^{4-\epsilon}}\frac{\Gamma(3-\frac{\epsilon}{2})
\Gamma(\frac{\epsilon}{2})}{\Gamma(2-\frac{\epsilon}{2})\Gamma(3)}\frac{(\epsilon-2)^2}{4-\epsilon}\int_0^1dx
\int_0^1dy\ 
y\frac{[Q^2xy^2(1-x)]^{\frac{-\epsilon}{2}}}{\left[1+\frac{\lambda^2(1-y)}{Q^2xy^2(1-x)}\right]^{\frac{\epsilon}{2}}}\
.
\end{equation}
Using the Mellin-Barnes representation
\begin{equation}
\frac{1}{\left[1+\frac{\lambda^2(1-y)}{Q^2xy^2(1-x)}\right]^{\frac{\epsilon}{2}}}=\frac{1}{2i\pi}
\int_{c-i\infty}^{c+i\infty}ds\left(\frac{\lambda^2(1-y)}{Q^2xy^2(1-x)}\right)^{-s}\frac{\Gamma(s)
\Gamma\left(\frac{\epsilon}{2}-s\right)}{\Gamma\left(\frac{\epsilon}{2}\right)}\ ,
\end{equation}
with $c\equiv\Re(s)\in]0,\frac{\epsilon}{2}[$ (the $s$-complex plane fundamental strip of the
 Mellin-Barnes representation), we
then get
\begin{eqnarray}
I_1=\mu^\epsilon2\frac{i}{(2\sqrt{\pi})^{4-\epsilon}}&&\!\!\!\!\!\!\!\!\frac{\Gamma(3-\frac{\epsilon}{2})}
{\Gamma(2-\frac{\epsilon}{2})\Gamma(3)}\frac{(\epsilon-2)^2}{4-\epsilon}(Q^2)^{-\frac{\epsilon}{2}}\nonumber\\
&\times&\!\!\frac{1}{2i\pi}\int_{c-i\infty}^{c+i\infty}ds\left(\frac{\lambda^2}{Q^2}\right)^{-s}\frac{\pi}{\sin(\pi
s)}\frac{\Gamma\left(\frac{\epsilon}{2}-s\right)\Gamma^2\left(1-\frac{\epsilon}{2}+s\right)}
{\Gamma\left(3-\epsilon+s\right)}\label{I1}
\ ,
\end{eqnarray}
where we performed the parametric integrals, which did not modify the fundamental strip.

Similarly, using 
\begin{equation}
\frac{1}{\left[1+\frac{\lambda^2(1-y)}{Q^2xy^2(1-x)}\right]^{1+\frac{\epsilon}{2}}}=\frac{1}{2i\pi}
\int_{c-i\infty}^{c+i\infty}ds\left(\frac{\lambda^2(1-y)}{Q^2xy^2(1-x)}\right)^{-s}\frac{\Gamma(s)
\Gamma\left(1+\frac{\epsilon}{2}-s\right)}{\Gamma\left(1+\frac{\epsilon}{2}\right)}\
,
\end{equation}
where the fundamental strip is given in this case by $c\in]0,1+\frac{\epsilon}{2}[$, we have for the second integral
 in the bracket of (\ref{F1})
\begin{eqnarray}
I_2&\equiv& \mu^\epsilon2\frac{-i}{(2\sqrt{\pi})^{4-\epsilon}}\frac{\Gamma(1+\frac{\epsilon}{2})}{\Gamma(3)}\nonumber\\
&\times&\int_0^1dx\int_0^1dy\  y[Q^2xy^2(1-x)+\lambda^2(1-y)]^{-1-\frac{\epsilon}{2}}Q^2[2(1-y)+(x-1)xy^2(\epsilon-2)]\nonumber\\
&=& \mu^\epsilon2\frac{-i}{(2\sqrt{\pi})^{4-\epsilon}}\frac{\Gamma(1+\frac{\epsilon}{2})}{\Gamma(3)}\left\{2Q^2\int_0^1dx\int_0^1dy\  y(1-y)[Q^2xy^2(1-x)+\lambda^2(1-y)]^{-1-\frac{\epsilon}{2}}\right.\nonumber\\
&+&\left.(\epsilon-2)Q^2\int_0^1dx\int_0^1dy\  (x-1)xy^3[Q^2xy^2(1-x)+\lambda^2(1-y)]^{-1-\frac{\epsilon}{2}}\right\}\nonumber\\
&=&\mu^\epsilon2\frac{-i}{(2\sqrt{\pi})^{4-\epsilon}}\frac{\Gamma(1+\frac{\epsilon}{2})}
{\Gamma(3)}(Q^2)^{-\frac{\epsilon}{2}}\nonumber\\
&\times&\left(2\frac{1}{2i\pi}\int_{d-i\infty}^{d+i\infty}ds
\left(\frac{\lambda^2}{Q^2}\right)^{-s}\frac{\Gamma(s)\Gamma\left(1+\frac{\epsilon}{2}-s\right)}
{\Gamma\left(1+\frac{\epsilon}{2}\right)}\frac{\Gamma^2\left(s-\frac{\epsilon}{2}\right)
\Gamma(2-s)}{\Gamma\left(s+2-\epsilon\right)}\right.\nonumber\\
\label{I2}
&-&\left.(\epsilon-2)\frac{1}{2i\pi}\int_{f-i\infty}^{f+i\infty}ds\left(\frac{\lambda^2}{Q^2}\right)^{-s}
\frac{\Gamma(s)\Gamma\left(1+\frac{\epsilon}{2}-s\right)}{\Gamma\left(1+\frac{\epsilon}{2}\right)}
\frac{\Gamma^2\left(1-\frac{\epsilon}{2}+s\right)\Gamma(1-s)}{\Gamma\left(s+3-\epsilon\right)}\right).
\end{eqnarray}
Notice that for the two Mellin-Barnes integrals in the last equation, the fundamental strips have been modified by 
the parametric integrations, since $d\in]\frac{\epsilon}{2},1+\frac{\epsilon}{2}[$ and $f\in]0,1[$.

It is possible to compute the asymptotic expansion of (\ref{I1}) and (\ref{I2}) in the $\lambda\rightarrow 0$ limit 
keeping an exact dependance in $\epsilon$, but we would get results on which the $\epsilon$ Laurent expansion would
be hard to obtain. In fact both integrals in (\ref{I2}) are convergent in the $\epsilon\rightarrow 0$ limit,
 which can then be performed at the integrand level in these integrals to give
\begin{equation}
\left.I_2\right\vert_{\epsilon\rightarrow0} =-\frac{i}{8\pi^2}\frac{1}{2i\pi}\int_{c-i\infty}^{c+i\infty}ds
\left(\frac{\lambda^2}{Q^2}\right)^{-s}\left(\frac{\pi}{\sin(\pi s)}\right)^2\frac{2-s}{s(s+1)(s+2)}\ ,
\end{equation}
where the fundamental strip is now $c\in]0,1[$.

\textit{A contrario}, the $\epsilon\rightarrow 0$ limit of (\ref{I1}) is not well-defined  since we have a pinch
 singularity in this limit (due to the fundamental strip $]0,\frac{\epsilon}{2}[$). This, of course, reflects the UV
divergence of the form factor. 

What we therefore do is to keep the exact $\epsilon$-dependance to compute the first term of the
 $\lambda\rightarrow 0$ asymptotic expansion of (\ref{I1}). After that, the pinch singularity being discarded,
the $\epsilon$-expansion becomes possible in the remainder integral directly at the integrand level. We then perform
renormalization, subtracting to the form factor its value at $Q^2=0$. 

To compute the first term of the $\lambda\rightarrow 0$ asymptotic expansion of (\ref{I1}), we follow \cite{FlaGoDu95,Friot:2005cu} (see also Appendix A).
One thus needs the first singular element to the left of the fundamental strip of (\ref{I1}),
 which is located at $s=0$ and reads 
\begin{equation}
\left.\left[\frac{\pi}{\sin(\pi s)}\frac{\Gamma\left(\frac{\epsilon}{2}-s\right)\Gamma^2
\left(1-\frac{\epsilon}{2}+s\right)}{\Gamma\left(3-\epsilon+s\right)}\right]\right\vert^{sing.}_{s\rightarrow
0}=\frac{1}{s}\frac{\Gamma\left(\frac{\epsilon}{2}\right)\Gamma^2
\left(1-\frac{\epsilon}{2}\right)}{\Gamma\left(3-\epsilon\right)}\
.
\end{equation}
We therefore conclude that
\begin{eqnarray}
I_1&=&\mu^\epsilon2\frac{i}{(2\sqrt{\pi})^{4-\epsilon}}\frac{\Gamma(3-\frac{\epsilon}{2})}
{\Gamma(2-\frac{\epsilon}{2})\Gamma(3)}\frac{(\epsilon-2)^2}{4-\epsilon}(Q^2)^{-\frac{\epsilon}{2}}
\frac{\Gamma\left(\frac{\epsilon}{2}\right)\Gamma^2\left(1-\frac{\epsilon}{2}\right)}{\Gamma\left(3-\epsilon\right)}\nonumber\\
&+&\ \mu^\epsilon2\frac{i}{(2\sqrt{\pi})^{4-\epsilon}}\frac{\Gamma(3-\frac{\epsilon}{2})}
{\Gamma(2-\frac{\epsilon}{2})\Gamma(3)}\frac{(\epsilon-2)^2}{4-\epsilon}(Q^2)^{-\frac{\epsilon}{2}}\nonumber\\
&&\ \ \ \ \ \ \ \ \ \ \ \times
\frac{1}{2i\pi}\int_{d-i\infty}^{d+i\infty}ds\left(\frac{\lambda^2}{Q^2}\right)^{-s}\frac{\pi}{\sin(\pi
s)}\frac{\Gamma\left(\frac{\epsilon}{2}-s\right)\Gamma^2\left(1-\frac{\epsilon}{2}+s\right)}
{\Gamma\left(3-\epsilon+s\right)}\ ,
\end{eqnarray}
where now $d\in]-1+\frac{\epsilon}{2},0[$, so that one can safely perform the $\epsilon\rightarrow 0$
 limit inside the integral. Before doing this, let us come back to eq.(\ref{F1}) to compute the contribution of
renormalization.
\begin{eqnarray}
\mathcal{F}_1(0,\epsilon)&=&(-i)\alpha_s4\pi
C_F2\frac{i}{(2\sqrt{\pi})^{4-\epsilon}}\mu^\epsilon\frac{\Gamma(3-\frac{\epsilon}{2})\Gamma(\frac{\epsilon}{2})}
{\Gamma(2-\frac{\epsilon}{2})\Gamma(3)}\frac{(\epsilon-2)^2}{4-\epsilon}\int_0^1dx\int_0^1dy\ 
y[\lambda^2(1-y)]^{\frac{-\epsilon}{2}}\nonumber\\
&=&(-i)\alpha_s4\pi
C_F2\frac{i}{(2\sqrt{\pi})^{4-\epsilon}}\mu^\epsilon\frac{\Gamma(3-\frac{\epsilon}{2})\Gamma(\frac{\epsilon}{2})}
{\Gamma(2-\frac{\epsilon}{2})\Gamma(3)}\frac{4(2-\epsilon)}{(4-\epsilon)^2}\lambda^{-\epsilon}\
.
\end{eqnarray}
One therefore finds
\begin{eqnarray}
\mathcal{F}_{1,R}(Q^2)&\equiv&\left.\left(\mathcal{F}_1(Q^2,\epsilon)-\mathcal{F}_1(0,\epsilon)\right)\right\vert_{\epsilon=0}\nonumber\\
&=&\left.(-i)\alpha_s4\pi
C_F\left(I_1+I_2-2\frac{i}{(2\sqrt{\pi})^{4-\epsilon}}\mu^\epsilon\frac{\Gamma(3-\frac{\epsilon}{2})
\Gamma(\frac{\epsilon}{2})}{\Gamma(2-\frac{\epsilon}{2})\Gamma(3)}\frac{4(2-\epsilon)}{(4-\epsilon)^2}
\lambda^{-\epsilon}\right)\right\vert_{\epsilon=0}\nonumber\\
&=&(-i)\alpha_s4\pi
C_F\left\{\frac{i}{32\pi^2}\left[3+2\ln\left(\frac{\lambda^2}{Q^2}\right)\right]\right.\nonumber\\
&&\ \ \ \ \ \ \ \  -\left.
\frac{i}{8\pi^2}\frac{1}{2i\pi}\int_{d-i\infty}^{d+i\infty}ds\left(\frac{\lambda^2}{Q^2}\right)^{-s}\left(\frac{\pi}{\sin(\pi
s)}\right)^2\frac{1}{(2+s)(1+s)}\right.\nonumber\\
\label{epszero}
&&\ \ \ \ \ \ \ \  -\left.\frac{i}{8\pi^2}\frac{1}{2i\pi}\int_{c-i\infty}^{c+i\infty}ds\left(\frac{\lambda^2}{Q^2}\right)^{-s}
\left(\frac{\pi}{\sin(\pi s)}\right)^2\frac{2-s}{(2+s)(1+s)s}\right\},
\end{eqnarray}
with $d\in]-1,0[$ and $c\in]0,1[$.
To get the complete $\lambda\rightarrow 0$ asymptotic expansions of the two integrals in the right-hand side of
(\ref{epszero}), we then have to compute the singular expansion\footnote{The singular expansion is simply the formal sum of all singular elements, and it is denoted by the symbol $\asymp$ as in \cite{FlaGoDu95,Friot:2005cu}.} of the Mellin transforms
\begin{equation}\label{M1}
\mathcal{M}_1(s)\equiv\left(\frac{\pi}{\sin(\pi s)}\right)^2\frac{1}{(2+s)(1+s)}
\end{equation}
and
\begin{equation}\label{M2}
\mathcal{M}_2(s)\equiv\left(\frac{\pi}{\sin(\pi s)}\right)^2\frac{2-s}{s(s+1)(s+2)}
\end{equation}
to the left of their corresponding fundamental strips (\textit{i.e} $]-1,0[$ for $\mathcal{M}_1$ and $]0,1[$ for
 $\mathcal{M}_2$).

One finds
$$
\mathcal{M}_1(s)\asymp \frac{1}{(s+1)^3}-\frac{1}{(s+1)^2}+\frac{1+\frac{\pi^2}{3}}{s+1}-\frac{1}{(s+2)^3}
-\frac{1}{(s+2)^2}-\frac{1+\frac{\pi^2}{3}}{s+2}
$$
\begin{equation}
\ \ \ \ \ \ \ +\sum_{n=3}^{\infty}\frac{1}{(2-n)(1-n)}\frac{1}{(s+n)^2}-\sum_{n=3}^{\infty}\frac{3-2n}{(2-n)^2(1-n)^2}
\frac{1}{s+n}
\end{equation}
and
$$
\mathcal{M}_2(s)\asymp \frac{1}{s^3}-\frac{2}{s^2}+\frac{\frac{5}{2}+\frac{\pi^2}{3}}{s}-
\frac{3}{(s+1)^3}+\frac{1}{(s+1)^2}-\frac{3+\pi^2}{s+1}+\frac{2}{(s+2)^3}+\frac{5}{2}\frac{1}{(s+2)^2}
+\frac{\frac{11}{4}+\frac{2\pi^2}{3}}{s+2}
$$
\begin{equation}
-\sum_{n=3}^{\infty}\frac{2+n}{(2-n)(1-n)n}\frac{1}{(s+n)^2}+\sum_{n=3}^{\infty}
\frac{-4+12n-3n^2-2n^3}{(2-n)^2(1-n)^2n^2}\frac{1}{s+n}\ .
\end{equation}
Now, since our Mellin transforms (\ref{M1}) and (\ref{M2}) fulfil the necessary condition of decrease along vertical lines \cite{FlaGoDu95}, we can apply the \textit{converse mapping theorem} \cite{FlaGoDu95,Friot:2005cu}, which gives the complete asymptotic expansion for
 $\lambda^2/Q^2\rightarrow 0$
\begin{eqnarray}
\mathcal{F}_{1,R}(Q^2)&\sim&(-i)\alpha_s4\pi
C_F\left\{\frac{i}{32\pi^2}\left[3+2\ln\left(\frac{\lambda^2}{Q^2}\right)\right]-\frac{i}{8\pi^2}
\left[\frac{1}{2}\frac{\lambda^2}{Q^2}\ln^2\left(\frac{\lambda^2}{Q^2}\right)+\frac{\lambda^2}{Q^2}
\ln\left(\frac{\lambda^2}{Q^2}\right)\right.\right.\nonumber\\
&+&\left(1+\frac{\pi^2}{3}\right)\frac{\lambda^2}{Q^2}
-\frac{1}{2}\left(\frac{\lambda^2}{Q^2}\right)^2\ln^2\left(\frac{\lambda^2}{Q^2}\right)
+\left(\frac{\lambda^2}{Q^2}\right)^2\ln\left(\frac{\lambda^2}{Q^2}\right)
-\left(1+\frac{\pi}{3}\right)\left(\frac{\lambda^2}{Q^2}\right)^2\nonumber\\
&-&\label{Asympt}\ln\left(\frac{\lambda^2}{Q^2}\right)
\sum_{n=3}^{\infty}\frac{1}{(2-n)(1-n)}\left(\frac{\lambda^2}{Q^2}\right)^n
\left.+\sum_{n=3}^{\infty}\frac{2n-3}{(2-n)^2(1-n)^2}\left(\frac{\lambda^2}{Q^2}\right)^n\right]\\
&-&\frac{i}{8\pi^2}\left[\frac{1}{2}\ln^2\left(\frac{\lambda^2}{Q^2}\right)+2\ln\left(\frac{\lambda^2}{Q^2}\right)
+\frac{5}{2}+\frac{\pi^2}{3}-\frac{3}{2}\frac{\lambda^2}{Q^2}\ln^2\left(\frac{\lambda^2}{Q^2}\right)\right.
\left.-\frac{\lambda^2}{Q^2}\ln\left(\frac{\lambda^2}{Q^2}\right)\right.\nonumber\\
&-&(3+\pi^2)\frac{\lambda^2}{Q^2}+\left.\left(\frac{\lambda^2}{Q^2}\right)^2\ln^2\left(\frac{\lambda^2}{Q^2}\right)
-\frac{5}{2}\left(\frac{\lambda^2}{Q^2}\right)^2\ln\left(\frac{\lambda^2}{Q^2}\right)
+\left(\frac{11}{4}+\frac{2\pi^2}{3}\right)\left(\frac{\lambda^2}{Q^2}\right)^2\right.\nonumber\\
&+&\left.\left.\ln\left(\frac{\lambda^2}{Q^2}\right)\sum_{n=3}^{\infty}\frac{2+n}{(2-n)(1-n)n}
\left(\frac{\lambda^2}{Q^2}\right)^n+\sum_{n=3}^{\infty}\frac{-4+12n-3n^2-2n^3}{(2-n)^2(1-n)^2n^2}
\left(\frac{\lambda^2}{Q^2}\right)^n\right]\right\}\nonumber
.
\end{eqnarray}
The first few terms are ($a_s\equiv\alpha_s/4\pi$)
\begin{eqnarray}
\mathcal{F}_{1,R}(Q^2)\sim &-&a_sC_F\left[\ln^2\left(\frac{\lambda^2}{Q^2}\right)+3\ln\left(\frac{\lambda^2}{Q^2}\right)
+\frac{7}{2}+\frac{2\pi^2}{3}\right.\nonumber\\
&&\ \ \ \ \ \ \ \ \ \ \ \ \ \ \ \ -2\left.
\left(\frac{\lambda^2}{Q^2}\right)\left(\ln^2\left(\frac{\lambda^2}{Q^2}\right)
+2+\frac{2\pi^2}{3}\right)+...\right]\ ,
\end{eqnarray}
where the non-analytic logarithmic  term in the correction which vanishes for $\lambda\rightarrow 0$ signals
 \cite{Ball:1995ni} the leading renormalon in the quark form factor. 
 
It is easy to prove that the asymptotic
expansion (\ref{Asympt}) is in fact an \textit{exact} result since all sums in (\ref{Asympt}) are convergent in our
limit (notice that they can be easily expressed in terms of usual functions after decomposition into partial fractions) and because there are no exponentially
suppressed terms. 

\noindent Indeed, one finds
\begin{equation}
\sum_{n=3}^{\infty}\frac{-4+12n-3n^2-2n^3}{(2-n)^2(1-n)^2n^2}\left(\frac{\lambda^2}{Q^2}\right)^n=
\frac{\lambda^2}{Q^2}-\frac{11}{4}\left(\frac{\lambda^2}{Q^2}\right)^2+\left[-1+
\left(3-2\frac{\lambda^2}{Q^2}\right)\frac{\lambda^2}{Q^2}\right]\textrm{Li}_2\left(\frac{\lambda^2}{Q^2}\right)\ 
\end{equation}
and similar results for the other sums in (\ref{Asympt}).

\noindent Moreover the absence of exponentially suppressed terms is due to the fact that the asymptotic remainder
 integrals tend to zero. 
Indeed, choosing $T=\frac{2j+1}{2}$ where $j\in \mathbb{N}$, we have\footnote{This inequality follows from the fact that $\displaystyle{\left|\int_\mathcal{C}ds f(s)\right|\leq ML}$, where $M$ is the maximum modulus of $f(s)$ on $\mathcal{C}$ and $L$ is the length of $\mathcal{C}$.}
\begin{equation}
\left\vert\int_{-T-iT}^{-T+iT}ds
\left(\frac{\lambda^2}{Q^2}\right)^{-s}\left(\frac{\pi}{\sin(\pi
s)}\right)^2\frac{1}{(2+s)(1+s)}\right\vert\leq2T\left\vert
\frac{\lambda^2}{Q^2}\right\vert^{T}\pi^2\left\vert\frac{1}{(2-T)(1-T)}\right\vert\ ,
\end{equation}
and the righthand side vanishes for $T\rightarrow +\infty$ if $\frac{\lambda^2}{Q^2}<1$. A similar result for the
other integral of (\ref{epszero}) is easily obtained.

After simplification of (\ref{Asympt}), one then has, because of the absence of exponentially suppressed terms, 
the final exact
 result 
\begin{eqnarray}
\mathcal{F}_{1,R}(Q^2)&=&a_sC_F\left\{\left(1-\frac{\lambda^2}{Q^2}\right)^2\left[2\ \textrm{Li}
_2\left(\frac{\lambda^2}{Q^2}\right)+2\ln\left(\frac{\lambda^2}{Q^2}\right)\ln\left(1-\frac{\lambda^2}{Q^2}\right)
-\ln^2\left(\frac{\lambda^2}{Q^2}\right)-\frac{2\pi^2}{3}\right]\right.\nonumber\\
\label{result2}
&&\left.\ \ \ \ \ \ \ \ \  -\frac{7}{2}+2\frac{\lambda^2}{Q^2}-\ln\left(\frac{\lambda^2}{Q^2}\right)\left[3-2\frac{\lambda^2}{Q^2}\right]\right\},
\end{eqnarray}
which is indeed equal to $a_sC_F2{\cal V}_s(\lambda^2/Q^2)= C_F{\cal V}_s(\lambda^2/Q^2)
\frac{\alpha_s}{2\pi}$  (eq.(\ref{result1})),
 with ${\cal V}_s(\lambda^2/Q^2)$ as defined in eq.(\ref{nu}).  Our result agrees with \cite{Dokshitzer:1995qm},
 but only provided we interpret
their ``total correction to the renormalized hard vertex" as \textit{twice} the one-loop renormalized quark form
factor, since the normalization of ${\cal V}_s(\lambda^2/Q^2)$ in this latter reference is twice the one used 
in eq.(\ref{nu}). This factor of 2 arises because it is the {\em square} of the form factor which occurs in
eq.(\ref{identity-0}).\\

\noindent
{\bf Note added in proofs}: the referee has pointed out to us that the all orders validity of eq. (2.37), hence of the conjecture eq. (2.26) in the DIS case, can actually be derived from the results of section 4 in the paper JHEP01(2007)076 by Becher, Neubert and Pecjak, once one notices that the matching function $C_V(Q^2,\mu)$ in this reference is related to $G(Q^2/\mu^2,a_s)$ by $G(Q^2/\mu^2,a_s)=\frac{d}{d\ln Q}\ln C_V(Q^2,\mu)$. We thank the referee for this valuable information.

\end{document}